\documentclass[twocolumn,prd,aps,amsmath,floatfix]{revtex4} 

\usepackage{epsfig} 
\newcommand{\bea}{\begin{eqnarray}} 
\newcommand{\eea}{\end{eqnarray}} 
\newcommand{\Bx}{x_{bj}}  
\font\cmss=cmss12  
\def\1{\hbox{{1}\kern-.25em\hbox{l}}} 
\def\bfZ{\relax{\hbox{\cmss Z\kern-.4em Z}}} 
 
\begin{document} 
\def\lsim{\mathrel{\rlap{\lower4pt\hbox{\hskip1pt$\sim$}}
    \raise1pt\hbox{$<$}}}                
\def\gsim{\mathrel{\rlap{\lower4pt\hbox{\hskip1pt$\sim$}}
    \raise1pt\hbox{$>$}}}                

\title{A detailed QCD analysis of twist-3 effects in DVCS observables} 
\author{A.~Freund}
\affiliation{Institut f{\"u}r Theoretische Physik, Universit{\"a}t Regensburg,  
D-93040 Regensburg, Germany} 
\medskip 

\begin{abstract} 
  In this paper I present a detailed QCD analysis of twist-3 effects
  in the Wandzura-Wilczek (WW) approximation in deeply virtual Compton
  scattering (DVCS) observables for various kinematical settings,
  representing the HERA , HERMES, CLAS and the planned EIC
  (electron-ion-collider) experiments. I find that the twist-3 effects
  in the WW approximation are almost always negligible at collider
  energies but can be large for low $Q^2$ and smaller $\Bx$ in
  observables for the lower energy, fixed target experiments directly
  sensitive to the real part of DVCS amplitudes like the charge
  asymmetry (CA).  Conclusions are then drawn about the reliability of
  extracting twist-2 generalized parton distributions (GPDs) from
  experimental data and a first, phenomenological, parameterization of
  the LO and NLO twist-2 GPD $H$, describing all the currently
  available DVCS data within the experimental errors is given.
\end{abstract} 
\maketitle 

\section{Introduction} 
 
Hard, exclusive processes and amongst them deeply virtual Compton
scattering (DVCS) in particular
\cite{mrgdh,ji,rad1,die,vgg,jcaf,ffgs,ffs,gb1,gb2,bemuothers}, have
emerged in recent years as prime candidates to gain a three
dimensional \cite{foot1} image of parton correlations inside a nucleon
\cite{afgpd3d,burk12,diehl3d,bemunew,ji3dnew}. This information is
gained by mapping out the key component containing this three
dimensional information, namely generalized parton distributions
(GPDs).

GPDs have been studied extensively in recent years
\cite{mrgdh,ji,rad1,die,vgg,jcaf,ffgs,ffs,gb1,gb2,bemuothers,shupoly}
since these distributions are not only the basic, non-perturbative
ingredient in hard, exclusive processes such as DVCS or exclusive
vector meson production, they are generalizations of the well known
parton distribution functions (PDFs) from inclusive reactions. GPDs
incorporate both a partonic and distributional amplitude behavior and
hence contain more information about the hadronic degrees of freedom
than PDFs. In fact, GPDs are true two-parton correlation functions,
allowing access to highly non-trivial parton correlations inside
hadrons \cite{foot}.

In order to perform a mapping of GPDs in their variables, experimental
data from a wide variety of hard, exclusive processes is needed.
Furthermore, in order to be able to properly interpret the data, the
processes should be well understood theoretically. Therefore, one
should start out exploring the theoretically ``simplest'' process i.e.
the one with the least theoretical uncertainty. In the class of hard,
exclusive processes this is DVCS ($e(k) + p(P) \to e(k') + p(P') +
\gamma(q')$). The reason for this is the simple structure of its
factorization theorem \cite{ji,rad1,jcaf}.  The scattering amplitude
is simply given by the convolution of a hard scattering coefficient
computable to all orders in perturbation theory with ${\it one}$ type
of GPD carrying the non-perturbative information. Factorization
theorems for other hard exclusive processes are usually ${\it double}$
convolutions containing more than one non-perturbative function.

As with all factorization theorems, the DVCS scattering amplitude is
given in this simple convolution form up to terms which are suppressed
in the large scale of the process. In this case the large scale is the
transfered momentum i.e. the virtual photon momentum, $Q$, and the
suppressed terms are of the order
$O((m_N/Q)^n),O((\sqrt{-t}/Q)^n,O((\lambda_{QCD}/Q)^n)$ with $m_N$ the
proton mass and $t = (P-P')^2$ the momentum transfer onto the outgoing
proton. This means, however, that for smaller values of $Q$ these
uncontrolled terms, called higher twist corrections, could in
principle be sizeable and the convolution or lowest twist term need
not be the leading one.  Since we are interested in the extraction of
the non-perturbative information of the lowest twist GPD, in this case
twist-2, we need to understand something about these higher twist
corrections.

It was shown by several groups
\cite{bemutw3,radweitw3,antertw3,kivpoly} that the first suppressed
term (twist-3) in the DVCS scattering amplitude can, in leading order
of the strong coupling constant (LO) and in the Wandzura-Wilczek (WW)
approximation, be simply expressed through a sum of terms which can be
written as a convolution of a LO twist-3 hard scattering coefficient
with a twist-2 GPD. Unfortunately, it could not be shown that twist-3
in the WW approximation does factorize to all orders in perturbation
theory (it seems to hold in NLO though \cite{kivmach}). However, we do
have at least some control over the leading of the higher twist terms
and can therefore use parameterizations of twist-2 GPDs also for
twist-3.

Equipped with this knowledge one might think that this information is
enough to obtain an unambiguous interpretation of DVCS data in terms
of GPDs, but life is unfortunately even more complicated. Besides the
dynamical twist contributions to the amplitude, there are kinematical
power corrections in the DVCS cross section i.e.
contributions where a dynamical twist-2 amplitude is multiplied by a
term $\propto(\sqrt{-t}/Q,m_N/Q)$ which makes them of the same power in
$1/Q$ as twist-3. These terms are of particular importance
in the interference term between DVCS and the QED Bethe-Heitler (BH)
process which both contribute to the total DVCS cross section. Since
the kinematical power corrections are nothing but dynamical twist-2 with a
kinematical dressing they can be handled by the same GPD
parameterization as dynamical twist-2. Thus we have the leading
corrections to the DVCS process under control, at least in LO, and can
now investigate several things: a) How big are physical observables at
various values of $x_{bj},Q^2$ and $t$ for a certain center-of-mass
energy $s$, especially those terms isolating various parts in the
interference term directly proportional to real or imaginary parts of
DVCS amplitudes? b) How big are the next-to-leading order (NLO)
corrections in the strong coupling constant to these observables? c)
How big are the corrections to these observables due to
twist-3 effects? d) How reliably can the twist-2 GPDs be extracted
from DVCS data?

In this paper I will concentrate on c) and d). a) and b) have been
extensively discussed using various GPD models in
\cite{afmmlong,afmmshort,bemu3}. c) and d) have, in various forms,
been discussed in \cite{bemu4,vanderhagen1}. This was done, however,
without taking evolution effects into account. We know that evolution
effects are sizeable and can change the shape of the GPD in the ERBL
region substantially \cite{afmmgpd}, thereby strongly affecting the
real part of the amplitude \cite{afmmamp} and the observables
associated with the real part. In \cite{afmmlong,afmmshort} it was
demonstrated that the type of GPD models most commonly used cannot be
brought into agreement with the available data once evolution effects
were taken into account. Subsequently, in \cite{afmmms}, a model was
proposed which was able to describe all currently available DVCS data
within a full NLO QCD analysis. 

I will extend the analysis of \cite{afmmms}, by not only using the
input model of \cite{afmmms} for the twist-2 GPDs but also as an input
for the twist-3 sector in the WW approximation.
Sec.~\ref{dvcsobs} contains the DVCS kinematics, structure of the
cross section, equations for the twist-3 contributions and the
definition of the relevant observables. In Sec.~\ref{gpdmodel}, I will
recapitulate the model I use for the four relevant twist-2 GPDs
$H,\tilde H, E, \tilde E$ and their twist-3 counter parts.
Sec.~\ref{resultsob} contains the twist-2 vs. twist-3 results for DVCS
observables in various kinematical settings. In Sec.~\ref{paramet}, I
will present for the first time a phenomenological parameterization of
the twist-2 GPD $H$ which can describe all currently available DVCS
data within the experimental errors in both LO and NLO QCD. I will
then summarize in Sec.~\ref{conc}.

\begin{figure} 
\centering 
\mbox{\epsfig{file=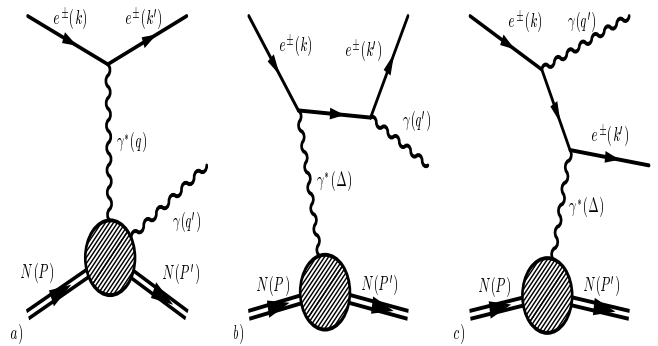,width=8.5cm,height=4.5cm}} 
\caption{a) DVCS graph, b) BH with photon from final state lepton and
  c) with photon from initial state lepton.}
\label{dvcspic} 
\end{figure}

\section{DVCS: Kinematics, cross section and definitions}
\label{dvcsobs}
The lepton level process, $e^{\pm} (k,\kappa) ~N(P,S) \to e^{\pm}(k',\kappa')
~N(P',S')~\gamma(q',\epsilon')$, receives contributions from each of the graphs
shown in Fig.\ \ref{dvcspic}. This means that the cross section will
contain a pure DVCS-, a pure BH- and an interference term.

\begin{figure}
\centering
\mbox{\epsfig{file=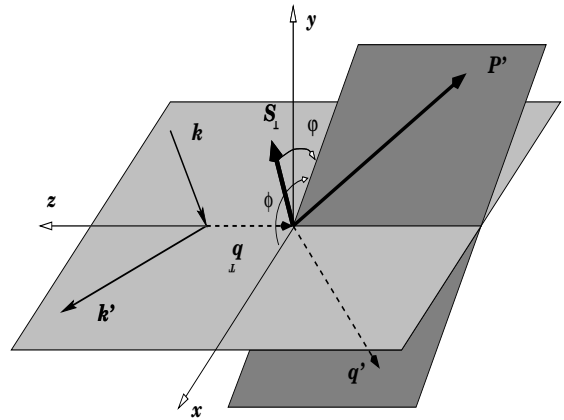,width=8.5cm,height=5.5cm}}
\caption{
  The kinematics of the leptoproduction in the target rest frame.}
\label{frame}
\end{figure}

I choose to work in the target rest frame given in \cite{bemu4} (see
Fig.~\ref{frame}), where the positive $z$-direction is chosen along
the three-momentum of the incoming virtual photon. The incoming and
outgoing lepton three-momenta form the lepton scattering plane, while
the final state proton and outgoing real photon define the hadron
scattering plane. In this reference frame the azimuthal angle of the
scattered lepton is $\phi_l = 0$, while the azimuthal angle between the
lepton plane and the final state proton momentum is $\phi_N = \phi$. When
the hadron is transversely polarized (within this frame of reference)
$S_\perp = (0, \cos {\mit\Phi}, \sin {\mit\Phi}, 0)$ and the angle between the
polarization vector and the scattered hadron is given by $\varphi = {\mit\Phi}
- \phi_N$. The four vectors are $k = (E, E \sin\theta_l, 0, E \cos\theta_l )$, $q =
(q^0, 0, 0,-|q^3|)$. Other vectors are $P = (M, 0, 0, 0)$ and $P' =
(E', |\mbox{\boldmath$P'$}| \cos\phi \sin\theta_N, | \mbox{\boldmath$P'$}|
\sin\phi \sin\theta_N, |\mbox{\boldmath$P'$}| \cos\theta_N)$. The longitudinal part
of the polarization vector is $S_{\rm LP} = (0, 0, 0, \Lambda)$. The
relevant Lorentz-invariant variables for DVCS are then:
\begin{eqnarray} 
\xi = \frac{Q^2}{2 {\bar P} \cdot {\bar q}} \, , 
{\bar {\cal Q}}^2 = -{\bar q}^2, 
t = \Delta^2 = (P-P')^2 \, ,
y = \frac{P\cdot q}{P\cdot k}\, , 
\nonumber 
\end{eqnarray} 
where ${\bar P} = (P+P')/2$, ${\bar q} = (q+q')/2$ and which are
related to the experimentally accessible variables, $\zeta \equiv x_{bj} =
-q^2/2P \cdot q$ and $Q^2 = - q^2$, used throughout this paper, via
\begin{align} 
&{\bar {\cal Q}}^2 = \frac{1}{2}Q^2\left(1+\frac{t}{Q^2}\right)\approx \frac{1}{2}Q^2\nonumber\\
&\xi = \frac{\zeta\left(1+\frac{t}{2 Q^2}\right)}{2-\zeta\left(1-\frac{t}{Q^2}\right)}\approx 
\frac{\zeta}{2-\zeta} \, . \label{qbar} 
\end{align}
Note that $t$ has a minimal value given by
\begin{eqnarray}
\label{Def-tmin}
-t_{\rm min}^2
=  Q^2
\frac{2(1 - \Bx) \left(1 - \sqrt{1 + \epsilon^2}\right) + \epsilon^2}
{4\Bx (1 - \Bx) + \epsilon^2}
\, .
\end{eqnarray}
where $\epsilon^2= 4M^2\Bx^2/Q^2$. Thus the theoretical limit $t\to0$ in an
exclusive quantity is not attainable in any experimental set-up and one
will have to rely on extrapolations.

The corresponding differential cross section is given by \cite{bemu4}: 
\begin{align} 
d\sigma = &\frac{(2\pi)^4}{4k \cdot P}|{\cal T}^{\pm}|^2 
\delta^{(4)} (k+P -k'-P'-q')\times\nonumber\\
& \frac{d^3{\bf k'}}{2k'_{0}(2\pi)^3}\frac{d^3{\bf P'}}{2 P'_{0} (2\pi)^3}\frac{d^3{\bf 
q'}}{2 q'_{0} (2\pi)^3} \, , 
\label{dvcscross} 
\end{align} 
and after integrating out some of the phase space we are left with a
five-fold differential cross section:
\begin{eqnarray}
\label{WQ}
\frac{d\sigma}{d\Bx dy d|t| d\phi d\varphi}
=
\frac{\alpha^3  \Bx y } {16 \, \pi^2 \,  Q^2 \sqrt{1 + \epsilon^2}}
\left| \frac{\cal T}{e^3} \right|^2 \, .
\end{eqnarray}

The square of the amplitude receives contributions from pure DVCS (Fig.\ 1a), from pure BH (Figs. 1b, 1c) and from their interference 
(with a sign governed by the lepton charge), 
\bea 
|{\cal T}|^2 = |{\cal T}_{DVCS}|^2 + {\cal I} + |{\cal T}_{BH}|^2   
\label{tdef} 
\eea 
where the individual terms are given by
\begin{eqnarray}
\label{Par-BH}
&&|{\cal T}_{\rm BH}|^2
= \frac{e^6}
{\Bx^2 y^2 (1 + \epsilon^2)^2 t\, {\cal P}_1 (\phi) {\cal P}_2 (\phi)}\times
\nonumber\\
&&\left[
c^{\rm BH}_0
+  \sum_{n = 1}^2
c^{\rm BH}_n \, \cos(n\phi) + s^{\rm BH}_1 \, \sin(\phi)
\right]
\\
\label{AmplitudesSquared}
&& |{\cal T}_{\rm DVCS}|^2
=
\frac{e^6}{y^2 {\cal Q}^2}\times\nonumber\\
&&\Big[
c^{\rm DVCS}_0
+ \sum_{n=1}^2
\left[c^{\rm DVCS}_n \cos (n\phi) + s^{\rm DVCS}_n \sin (n \phi)\right]
\Big]
\\
\label{InterferenceTerm}
&&{\cal I}
= \frac{\pm e^6}{\Bx y^3 t {\cal P}_1 (\phi) {\cal P}_2 (\phi)}\nonumber\\
&&\Big[
c_0^{\cal I}
+ \sum_{n = 1}^3
\left[c_n^{\cal I} \cos(n \phi) +  s_n^{\cal I} \sin(n \phi)\right]
\Big]
\end{eqnarray}
where the $+/-$ sign in the interference stands for a
negatively/positively charged lepton.

The $c_n$'s and $s_n$'s are the Fourier coefficients of the $\cos(n\phi)$
and $\sin(n\phi)$ terms. These coefficients are given as combinations of
the real and imaginary part of the unpolarized and the polarized
proton spin-non-flip and spin-flip DVCS amplitudes ${\cal
  H},{\cal\tilde H},{\cal E},{\cal\tilde E}$ (for the $c^{\cal I}$'s
or $s^{\cal I}$'s) or the squares of the afore mentioned DVCS
amplitudes (for the $c^{\rm DVCS}$'s or $s^{\rm DVCS}$'s). The exact
from is given in \cite{bemu4} and does not have to be repeated here. I
will discuss the computation of the DVCS amplitudes and the necessary
model assumptions in the next section. The precise form of the BH
propagators ${\cal P}_{1,2}(\phi)$ which induces an additional
$\phi$-dependence, besides the $\cos(n\phi)$ and $\sin(n\phi)$ terms, and which
can mock $\cos(n\phi)$ and $\sin(n\phi)$ dependences in certain observables,
can also be found in \cite{bemu4}.  Note that in order to avoid
collinear singularities occurring through the incidence of the
outgoing photon with the incoming lepton line in ${\cal P}_{1,2}(\phi)$
we need to constrain $y$ according to \bea y \leq y_{\rm col} \equiv \frac{Q^2
  + t}{Q^2 + \Bx t}.  \eea in order to avoid an artificially enhanced
BH contribution. This limit is only of practical relevance for fixed
target experiments at very low energies. Collider experiments do not
have any meaningful statistics for exclusive processes at very large
$y\simeq O(1)$. In the following discussion, I will neglect the
contributions to the DVCS cross section containing transversity and
terms higher than twist-3.

The DVCS observables I will deal with later on are based on a less
differential cross section than the five-fold one in Eq.~(\ref{WQ}). The
reason for this is first that the cross section in Eq.~(\ref{WQ}) is
frame dependent since the azimuthal angles $\phi$ and $\varphi$ are not Lorentz
invariants and hence, they will be integrated out. Secondly, since a
$t$-distribution is notoriously hard to measure, we also integrate out
$t$, however with experimentally sensible cuts as will be discussed
later. In consequence, our observables will be based on only a
two-fold differential cross section. Note that the DVCS data currently
available is at most for two-fold quantities, normally just one-fold
or even totally integrated over. One might argue that the more
variables in an observable are integrated out the more information is
lost, especially when studying higher twist effects. This is indeed
true, however, one has to make a sensible compromise between wishful
thinking on the one hand and experimental facts on the other. Also, as
I will show below, these two-fold quantities are enough to clearly
demonstrate the size of the higher twist effects on DVCS observables.
In the following, I will concentrate both for the sake of brevity and
the fact that these quantities are the easiest once to study, on the
Single Spin Asymmetry (SSA) and the Charge Asymmetry (CA) defined in
accordance with experiments, the following way:
\begin{align}
&SSA = \frac{2 \int_0^{2\pi}d\phi ~\sin(\phi)(d\sigma^{\uparrow}-d\sigma^{\downarrow})}{\int_0^{2\pi}d\phi~(d\sigma^{\uparrow}+d\sigma^{\downarrow})} \, ,
\label{defssa}\\
&CA =  \frac{2\int_0^{2\pi}d\phi ~\cos(\phi)(d\sigma^{+}-d\sigma^{-})}{\int_0^{2\pi}d\phi~(d\sigma^{+}+d\sigma^{-})} \, . 
\label{defca}
\end{align}
Here $d\sigma^{\uparrow}$ and $d\sigma^{\downarrow}$ refer to the two fold differential cross
sections $d\sigma/d\Bx dQ^2$ with the lepton polarized along or against its
direction of motion, respectively; $d\sigma^{+}$ and $d\sigma^{-}$ are the
unpolarized differential cross sections for positrons and electrons,
respectively.

Even though I am trying to discuss the charge asymmetry (CA) for two
experiments, EIC and CLAS, which cannot measure it at all since they
are or will be running with electrons only, there exist experimental
problems in measuring the proper quantity, the azimuthal angle
asymmetry or AAA. The AAA is defined below 
\bea 
\mbox{AAA} =\frac{\int^{\pi/2}_{-\pi/2} d\phi (d\sigma-d\sigma^{BH}) - \int^{3\pi/2}_{\pi/2}
  d\phi (d\sigma-d\sigma^{BH})}{\int^{2\pi}_{0} d\phi d\sigma}, \nonumber\\
\label{aaadef} 
\eea 

where $d\sigma^{BH}$ refers only to the pure BH cross section. The
experimental problem or challenge with the AAA is that it requires
either a very good detector resolution i.e. many bins in $\phi$ or an
event by event reconstruction of the scattering planes. The last
statement needs a word of explanation: Eq.~(\ref{aaadef}) is
equivalent to taking the difference between the number of DVCS minus
BH events where the real $\gamma$ is above the electron scattering plane
and where it is below that plane, divided by the total number of
events. This procedure ensures that the numerator is not contaminated
by BH, which would spoil an unambiguous interpretation of the
observable in terms of the real part of DVCS amplitudes. Also, the
only difference between Eq.~(\ref{defca}) and (\ref{aaadef}) is due to
the additional interference term in the denominator of
Eq.~(\ref{aaadef}), which is small compared to the leading
contribution, and a twist-2$\times$twist-3 term in
Eq.~(\ref{AmplitudesSquared}), which is both suppressed in the
kinematics considered and small in the employed  GPD model.

Unfortunately, in the case of CLAS where BH is by far the dominant
contribution in the cross section, one would need to subtract two
large numbers for the above plane and below plane events inducing a
huge statistical uncertainty. Therefore, I will not discuss the CA for
CLAS kinematics.

Let me say a word, about the expected effects of uncalculated higher
twist contributions besides the calculable twist-3 contributions.
Assume an asymmetry $A = \frac{B + C}{X + Y}$ where $B$ and $X$ stand
for leading twist contributions and and $C$ and $Y$ for the higher
twist contributions in the interference and the cross section term
respectively. Expanding the denominator yields $A= B/X + C/X - B\cdot Y/X^2
- C\cdot Y/X^2 + \ldots$. This shows that the leading higher twist contributions
in $A$ will originate from the twist corrections to the leading twist
interference term and will thus be of $O(\sqrt{-t}/Q,m_N/Q)$ with
respect to the leading term. Let us give a rough numerical example to
see the significance of the higher twist corrections: Assume
$B=1,X=5,C=\pm0.3$ and $Y=\pm1$, then $A=0.2$ in the leading twist
approximation and $A\simeq0.217$ in the full result. This shows that even
though we have $20-30\%$ higher twist effects in the individual terms,
they effectively cancel in the asymmetry if the corrections are both
positive. However, if one or both of the higher twist corrections are
negative, the result can vary between $O(20-40\%)$ from the leading
twist result. The relative sign of the higher tiwst terms in both
numerator and denominator will vary depending on the kinematic region
one is exploring and, therefore, it is a priori not clear what the
size of the higher twist corrections will be. Of course, these
arguments only hold if the higher twist contributions do not become of
order of the leading twist corrections.
  
In Sec.~\ref{paramet}, I will also talk about the one-photon cross
section $\sigma(\gamma^*p)$ at small $\Bx$ defined through
\begin{align} 
&\frac{d^{2}\sigma (ep \to ep\gamma)}{dy dQ^2} = \Gamma~\sigma_{DVCS}(\gamma^*p\to 
\gamma p)\nonumber\\
&~\qquad~\mbox{where}~~\qquad~\Gamma = \frac{\alpha_{e.m.} (1+(1-y)^2)}{2\pi 
  y Q^2}. 
\end{align}
with
\begin{align} 
\sigma_{DVCS}(\gamma^*p \to \gamma p) = \frac{\alpha^2x^2\pi}{Q^4{\cal B}}|{\cal 
  T}_{DVCS}|^2|_{t=0},
\label{sigonephot} 
\end{align} 
and where ${\cal B}$ stems from the $t$-integration and will depend on
both our cut-off in $t$ and the model of the $t$-dependence I will
choose for the GPDs. Furthermore, all higher twist effects are
neglected in this quantity.

\section{The GPD model and DVCS amplitudes: Twist-2 and Twist-3}
\label{gpdmodel}

\subsection{Modeling Twist-2 GPDs}
\label{modtw2}

In the following I will use and review the model for twist-2 GPDs
first introduced in \cite{afmmms}.

Based on the aligned jet model (AJM) (see for example \cite{fs88}) the
key Ansatz of \cite{afmmms} in the DGLAP region is:
\begin{equation}
H^{S,V,g} (X,\zeta) \equiv \frac{q^{S,V,g}\left(\frac{X-\zeta/2}{1-\zeta/2}\right)}{1-\zeta/2}\, ,
\label{fwd}
\end{equation}
where $q^i$ refers to any forward distribution and $H^{S,V} (X,\zeta)=
H^{q} (X,\zeta)\pm H^{\bar q} (X,\zeta)$. This Ansatz in the DGLAP region
corresponds to a double distribution model \cite{rad1,rad2,rad3} with an
extremal profile function allowing no additional skewdness save for
the kinematical one.It will also be used for $\tilde H$ and $E$.
I will talk more about the exact details for $\tilde H$ and $E$ below.
Note that I choose a GPD representation first introduced in
\cite{gb1}, which is maximally close to the inclusive case i.e
$X\in[0,1]$, $\zeta=\Bx$ with the partonic or DGLAP region in $[\zeta,1]$ and
the distributional ampitude or ERBL region in $[0,\zeta]$.

The prescription in Eq.~(\ref{fwd}) does not dictate what to do in the
ERBL region, which does not have a forward analog. The GPDs have to be
continuous through the point $X=\zeta$ and should have the correct
symmetries around the midpoint of the ERBL region. They are also
required to satisfy the requirements of polynomiality:
\begin{align}  
M_N & = \int^1_{\zeta} \frac{dX \tilde X^{N-1}}{2-\zeta}\Big[H^q (X,\zeta)-(-1)^{N-1}H^{\bar q}(X,\zeta)\nonumber\\
& + \frac{(1+(-1)^{N})}{2}\tilde X H^g (X,\zeta)\Big] \nonumber \\
& +(-1)^{N}\int^{\zeta}_0\frac{dX \tilde X^{N-1}}{2-\zeta}\left[H^{\bar q}(X,\zeta) +\tilde X H^g (X,\zeta)\right]\nonumber\\
     &=  \sum^{N/2}_{k=0} \left(\frac{\zeta}{2-\zeta}\right)^{2k} C_{2k,N} \, , 
\label{polynomiality}  
\end{align}
with $\tilde X = \frac{X-\zeta/2}{1-\zeta/2}$.  The ERBL region is therefore
modeled with these natural features in mind. One demands that the
resultant GPDs reproduce the first moment $M_1 = 3$ and the second
moment $M_2 = 1+C\zeta^2/(2-\zeta)^2$ \cite{footmom}.  $C$ was computed in the
chiral-quark-soliton model \cite{vanderhagen1} and found to be $-3.2$
and is related to the D-term \cite{poly} which lives exclusively in
the ERBL region.  This reasoning suggests the following simple
analytical form for the ERBL region ($X < \zeta$):
\begin{align}
&H^{g,V}(X,\zeta) = H^{g,V}(\zeta) \left[ 1+A^{g,V}(\zeta) C^{g,V} (X,\zeta) \right] \, , \nonumber \\
&H^{S}(X,\zeta) = H^{S}(\zeta)\left(\frac{X - \zeta/2}{\zeta/2}\right) \left[1 + A^{S}(\zeta) C^{S} (X,\zeta) \right] \, ,
\label{ajmerbl}
\end{align}
where the functions  
\begin{align}
&C^{g,V} (X,\zeta) = \frac{3}{2}\frac{2-\zeta}{\zeta}\left(1 - \left( \frac{X-\zeta/2}{\zeta/2} \right)^2 \right)  \, , \nonumber \\
&C^{S} (X,\zeta)  = \frac{15}{2}\left(\frac{2-\zeta}{\zeta}\right)^2 \left(1 - \left(\frac{X-\zeta/2}{\zeta/2} \right)^2 \right) \, ,
\label{defCs}
\end{align} 
vanish at $X=\zeta$ to guarantee continuity of the GPDs.  The $A^i(\zeta)$ are
then calculated for each $\zeta$ by demanding that the first two moments
of the GPDs are explicitly satisfied. For the second moment, what is
done in practice is to set the D-term to zero and demand that for each
flavor the whole integral over the GPD is equal to the whole integral
over the forward input PDF without the shift. For the final GPD, of
course, the D-term is added to the quark-singlet (there is no D-term
in the non-singlet sector) using the results from the chiral-quark-soliton
model \cite{vanderhagen1}. The gluonic D-term, about which nothing is
known save its symmetry, is set to zero for $Q_0$. Due to the
gluon-quark mixing in the singlet channel, there will be a gluonic
D-term generated through evolution. I will come back to this question
in Sec.~\ref{resultsob} when discussing DVCS for the HERMES
experimental setting.

It would be straightforward to extend this algorithm to satisfy
polynomiality to arbitrary accuracy by writing the $A^i(\zeta)$ explicitly
as a polynomial in $\zeta$ where the first few coefficients are set by the
first two moments and the other coefficients are then either
determined by the arbitrary functional form, as is done here, or,
perhaps theoretically more appealing, one chooses orthogonal
polynomials, such as Gegenbauer polynomials, for which one can set the
unknown higher moments equal zero. Phenomenologically speaking, the
difference between the two choices is negligible.

The above Ansatz also satisfies the required positivity conditions
\cite{rad2,posi1,posi2} and is in general extremely flexible both in
its implementation and adaption to either other forward PDFs or other
functional forms in the ERBL region. Therefore it can be easily
incorporated into a fitting procedure making it phenomenologically
very useful. In what follows we will use MRST2001 \cite{mrst01} and
CTEQ6 \cite{cteq6} as the forward distributions for both LO and NLO.

Let me quickly explain, why I only model certain C even and odd
distributions in the quark sector, $H^{S,V} (X,\zeta)= H^{q} (X,\zeta)\pm
H^{\bar q} (X,\zeta)$. As one can see below, only the quark charge
weighted $H^{S}$ appears in the DVCS amplitude and not $H^{V}$ due to
the C-even nature of the amplitude. The evolution equations for the
GPDs are defined for the following C-even and C-odd singlet (s) and
non-singlet (ns) flavor combinations:
\begin{align}
&H_{+}^{ns} = H^{q} + H^{{\bar q}}-\frac{1}{N_F}\sum_q(H^{q} + H^{\bar q})\, ,\nonumber\\
&H^{V}=H_{-}^{ns} = H^{q,} - H^{{\bar q}}\, ,\nonumber\\
&H^s=\frac{1}{N_F}\sum_q(H^{q} + H^{\bar q})=\frac{1}{N_F}\sum_qH^S\, ,
\label{quarkcomb1}
\end{align} 
where $H^s$ mixes, of course, with the gluon and $H_{\pm}^{ns}$ mix
neither with each other nor with the singlet and the gluon. A single
quark species, i.e. quark or anti-quark in the DGLAP region, or just a
singlet or non-singlet quark combination in the ERBL region (due to
the symmetry relations between quark and anti-quark in the ERBL region
in the off-diagonal representation of \cite{gb1}) can be extracted the
following way:
\begin{align} 
&\left(H^{q} \atop H^{\bar q}\right) = \frac{1}{2}(H_{+}^{NS} \pm H_{-}^{NS} + H^s) \,
\label{quarkcomb2}
\end{align} 
in the DGLAP region and 
\begin{align} 
&H^{S} = (H_{+}^{NS} + H^s) \, , \quad \mbox{and} \quad H^V = H_{-}^{ns}\, ,
\label{quarkcomb3}
\end{align} 
in the ERBL region. Eqs.~(\ref{quarkcomb1},\ref{quarkcomb2},\ref{quarkcomb3})
demonstrate that it is enough to model $H^{S,V}$ in order to
properly do evolution and extract the quark combinations relevant for DVCS.

The construction of $\tilde H$ proceeds analogous to that of $H$ with
opposite symmetries in the quark and gluon sector and using the
standard GRSV scenario \cite{grsv} as the forward input. Due to the
change of symmetry in the ERBL region the analytical form changes to
\begin{align}
&{\tilde H}^{S}(X,\zeta) = {\tilde H}^{S}(\zeta) \left[ 1+A^{S}(\zeta) C^{S} (X,\zeta) \right] \, , \nonumber \\
&{\tilde H}^{g,V}(X,\zeta) = {\tilde H}^{g,V}(\zeta)\left(\frac{X - \zeta/2}{\zeta/2}\right) \times\nonumber\\
&\left[1 + A^{g,V}(\zeta) C^{g,V} (X,\zeta) \right] \, ,
\label{ajmerblothersym}
\end{align}
\begin{align}
&C^{S} (X,\zeta) = \frac{3}{2}\frac{2-\zeta}{\zeta}\left(1 - \left( \frac{X-\zeta/2}{\zeta/2} \right)^2 \right)  \, , \nonumber \\
&C^{g,V} (X,\zeta)  = 4\frac{2-\zeta}{\zeta} \left(1 - \left(\frac{X-\zeta/2}{\zeta/2} \right)^2 \right) \, ,
\label{defCsothersym}
\end{align} 
Note that there is no D-term for the polarized GPDs due to symmetry
requirements.

For $\tilde E$, the asymptotic pion distribution amplitude i.e.  the
same Ansatz as in \cite{afmmms,bemu4} was used (see also references
therein for the same or similar Ans{\"a}tze).

The Ansatz for $E$ used in this paper deviates from the one used in
other studies to also include the gluon. First let me say a few
general things about $E$: The symmetries for $E$ are the same as the
ones for $\tilde H$ in the ERBL region. Furthermore, one would naively
expect that $E$ as a function of $\zeta=\Bx$ dies out as $\zeta$ decreases
i.e. behaves like a valence quark distribution. This is nothing but
the statement that it becomes increasingly difficult to flip the spin
of the proton as $\Bx$ is decreasing.  Also, we know that the first
moment of $E$ in the proton/neutron has to reproduce the respective
anomalous magnetic moments (see for example \cite{vanderhaeghen}).
This leads to

\bea
\kappa_u = 2\kappa_p + \kappa_n = 1.673\nonumber\\
\kappa_d = \kappa_p + 2\kappa_n = -2.033\, .  
\eea 

Following \cite{vanderhaeghen}, the ``forward'' quark distributions
from which to start is chosen to be
\begin{align}
&E^u(x) = \frac{1}{2}u_{val}(x)\cdot\kappa_u\nonumber\\
&E^d(x) = d_{val}(x)\cdot\kappa_d\nonumber\\
&E^s(x) = 0
\label{edefq}
\end{align}

For the DGLAP region and the quark singlet channel I will apply to
Eq.~(\ref{edefq}) the same shift as in Eq.~(\ref{fwd}). The sea
contribution is set to zero in the DGLAP region as was also done in
\cite{vanderhaeghen}. In contrast to \cite{vanderhaeghen}, I choose to
include a gluonic contribution. This contribution is important since
these distributions are evolved from a low scale $Q_0$ to the relevant
experimental scale in contrast to other groups
\cite{vanderhaeghen,bemu4} which chose to neglect evolution effects.
I model the gluon $E^g$ in the DGLAP region in the following way:

First, we know that the total angular momentum of the proton $J^p$ for
$\zeta=0$ and $t=0$ is given by:

\begin{align}
J^p= \frac{1}{2} = & \frac{1}{2}\int^1_0dX~\big[X(H^S(X) + E^S(X))\nonumber\\
& + H^g(X) + E^g(X)\Big]. 
\label{jatzeq0}
\end{align}
Since together with Eq.~(\ref{edefq}) I know now all the functions in
Eq.~(\ref{jatzeq0}) except the last, the contribution of the gluon to
$J^p$ can be defined to be $J^g_E = \frac{1}{2}(1 - J^S - J^g_H)$.
Thus one can define $E^g$ in the DGLAP region through
\begin{align}
E^g(X,\zeta) = \frac{J^g_E}{J^g_H}H^G(X,\zeta).
\label{edefg}
\end{align}
Note that in this particular model for the $E^q$, the second moment of
$E^S$ is zero due to symmetry, and since the second moment of the
$H^i$ already saturates the angular momentum sum rule, the gluon
contribution should be strictly zero. However, numerically, the second
moment of the $H^i$ is never exactly $1$, yielding a small but
non-zero $E^g$ and secondly, the above construction is general and
does not depend on the particulars of the model.

In the ERBL region, I will use the same strategy as for $H$ except
that I require the sum rule for $J^p$ to be fulfilled after the shift
in $X$ rather than the first and second moment of just the $H^i$.
Keeping the symmetry requirements for the $E^i$ in mind, this means
that one can use Eq.~(\ref{ajmerblothersym}) and (\ref{defCsothersym})
for the $E^i$ just with different $A^i(\zeta)$. In consequence, the
shifted version of Eq.~(\ref{edefq}) together with
Eqs.~(\ref{ajmerblothersym}), (\ref{defCsothersym}) and (\ref{edefg}),
gives a complete parameterization of the $E^i$. Note that this
parameterization of the $E^i$ is in stark contrast to \cite{bemu4}
where the $E^i$ are of the same size or even larger than the $H^i$
even at small $\Bx$ where this relation cannot hold. Hence the results
for the asymmetries in this paper will vary from those in \cite{bemu4}
at small $\Bx$ whereas in the valence region they will be similar.
Note furthermore that $E$ contains a D-term in both the quark-singlet
and the gluon. This D-term is identical to the one in $H$ but enters
with the opposite sign and thus cancels when considering the moments
of the sum of $H$ and $E$ as done for the total angular momentum
$J^p$.

As far as the $t$-dependence is concerned, I choose to model it the
same way as in \cite{afmmlong} i.e. using a factorized Ansatz for the
$t$-dependence from the rest of the GPD. I want to stress here that
this is not really realistic theoretical assumption especially at
larger values of $t$ \cite{vanderhaeghen}. On the experimental side,
it was shown in \cite{afmmms} that in order to describe the ZEUS data
on DVCS \cite{zeus} at large $Q^2$, a $Q^2$ dependent slope of the
$t$-dependence was required to describe the data. This in turn implies
that the basic assumption of a factorized $t$ dependence as well as
the assumption that the $t$ dependence of quarks and gluons is the
same is wrong, at least at large $Q^2$, since the $Q^2$ dependence of
the GPD can only be generated through perturbative evolution. The H1
data on DVCS \cite{h1}, for example, which lies in a lower $Q^2$ range
does not a priori require a $Q^2$ dependent slope.  This in turn means
that at low $Q^2$ and low $t$, where most of the experimental data
lies, a factorized $t$ dependence can still be used at the moment. The
situation improves even more when one considers asymmetries since
there either the $t$ dependence partially cancels between numerator
and denominator, if DVCS is dominant, or, if BH is dominant, the $Q^2$
range is such that a factorized approach is still not totally
unreasonable. As the accuracy and the kinematic reach of the data
improves, however, one has to seriously address the issue of a
non-factorized $t$-dependence of the GPD. I will discuss the issue in
detail in Sec.~\ref{paramet} and propose a phenomenological solution.

After having evolved the GPDs using the same program successfully
employed in \cite{afmmgpd}, the real and imaginary part of the twist-2
DVCS amplitude in LO and NLO given below are calculated using the same
program as in \cite{afmmamp}:
\begin{align}  
&{\cal T}^{S,V/A}_{DVCS} (\zeta,\mu^2,Q^2) = \sum_a e^2_a \left(\frac{2 - \zeta}{\zeta} \right)\times\nonumber\\
&\Big[  
\int^1_0 dX~T^{S_a,V/A} \left(\frac{2X}{\zeta} - 1+i\epsilon, \frac{Q^2}{\mu^2} \right) ~{\cal F}^{S_a,V/A} (X,\zeta,\mu^2)  \nonumber\\  
&\Big. \mp \int^1_{\zeta} dX~T^{S_a,V/A} \left(1 - \frac{2X}{\zeta},\frac{Q^2}{\mu^2}\right)~{\cal F}^{S_a/A} (X,\zeta,\mu^2) \Big] ,\nonumber\\
&{\cal T}^{g,V/A}_{DVCS} (\zeta,\mu^2,Q^2) = \frac{1}{N_f}\left (\frac{2 - \zeta}{\zeta}\right )^2 \times\nonumber\\
&\Big[ 
\int^1_0 dX~T^{g,V/A} \left(\frac{2X}{\zeta} - 1+i\epsilon, \frac{Q^2}{\mu^2} \right) ~{\cal F}^{g,V/A} (X,\zeta,\mu^2)  \nonumber\\
&\pm \Big. \int^1_{\zeta} dX~T^{g,V/A} \left(1 - \frac{2X}{\zeta}, \frac{Q^2}{\mu^2}\right) ~{\cal F}^{g,V/A}(X,\zeta,\mu^2) \Big] .  
\label{tdvcs}  
\end{align} 
$V/A$ stands for the vector/axial-vector i.e.
unpolarized/polarized case and ${\cal F}$ stands for the appropriate
GPD $H,\tilde H, E$ or $\tilde E$.
 
The $+i\epsilon$ prescription is implemented using the Cauchy principal value
prescription (``P.V.'') through the following algorithm:
\begin{align} 
&\Big. P.V. \int^1_0 dX~T\left(\frac{2X}{\zeta} - 1\right) {\cal F}(X,\zeta,Q^2) =\nonumber\\
&\int^{\zeta}_0 dX~T\left(\frac{2X}{\zeta} - 1\right)\left({\cal F}(X,\zeta,Q^2)-{\cal F}(\zeta,\zeta,Q^2)\right) + \nonumber\\ 
& \int^1_{\zeta} dX~T\left(\frac{2X}{\zeta} - 1\right) \left({\cal F}(X,\zeta,Q^2) -{\cal F}(\zeta,\zeta,Q^2)\right)\nonumber\\
&+{\cal F}(\zeta,\zeta,Q^2)\int^1_0 dX~T \left(\frac{2X}{\zeta} - 1\right) \, . 
\label{subtraction} 
\end{align} 
The relevant LO and NLO coefficient functions can be found in
\cite{bemu3,afmmamp}.

\subsection{Modeling Twist-3 GPDs}
\label{modtw-3} 

After having modeled the twist-2 sector which automatically takes care
of the kinematic power corrections in the DVCS cross section as
well, only the genuine twist-3 sector remains. As shown in
\cite{bemutw3,radweitw3,antertw3} the twist-3 GPDs and thus twist-3 DVCS
amplitudes can be expressed in the WW approximation through a
combination of twist-2 GPDs convoluted with a twist-3 coefficient
function.

The general structure of the twist-3 DVCS amplitudes can be found in
 Eq.~(84) of \cite{bemu4} and reads in the representation of \cite{gb1}:
\begin{align}
&{\cal T}^{tw-3}(\Bx,Q^2) = {\cal T}^{tw-2}(\Bx,Q^2)\nonumber\\
&+ \Bx\frac{\partial}{\partial\Bx} C^{V,A}_{tw-3}\otimes{\cal F} + \frac{2m_p^2\Bx}{(1-\Bx)(t-t_{min})}{\cal T}^{\perp}(\Bx,Q^2)\nonumber\\
& + {\cal T}^{qgq}(\Bx,Q^2)\, ,
\label{deftw-3}
\end{align}
where the first three terms are the WW terms and the last term is a
genuinely new dynamical contribution arising from $qgq$ correlations
which will be neglected in the following. The ${\cal
  T}^{\perp}(\Bx,Q^2)$'s are linear combinations of the type
$C^{V,A}_{tw-3}\otimes{\cal F}$ which can be found in Eq.~(87) of \cite{bemu4}
and do not have to be repeated here. Note that Eq.~(\ref{deftw-3})
differs from Eq.~(84) in \cite{bemu4} by a factor $(2-\Bx)$ which I have
pulled out for convenience. The convolution $C^{V,A}_{tw-3}\otimes{\cal F}$
is done using Eqs.~(\ref{tdvcs}) and (\ref{subtraction}) with the
$C^{V,A}_{tw-3}$ in LO having the following form
\begin{align}
C^{V,A}_{tw-3}(X,\zeta) = - \frac{\zeta}{2X}\ln\left(1-X/\zeta  +i\epsilon\right)\, .
\label{coefftw-3}
\end{align}
The respective subtraction factors $I^i$ i.e. $\int^1_0 dX~T
\left(\frac{2X}{\zeta} - 1\right)$, for the imaginary and real part of the
amplitude read
\begin{align}
&Re~I^{V,A}(\zeta) = \frac{\pi}{6}\zeta - \frac{1}{2}\zeta\mbox{Li}_2(\zeta) - \frac{1}{4}\zeta\ln(\zeta)^2\nonumber\\
&Im~I^{V,A}(\zeta) = - \frac{1}{2}\zeta\ln(\zeta)\, .  
\end{align}
${\cal T}^{tw-3}$ i.e. in particular the convolution and the
derivative of the convolution with respect to $\Bx$, was computed
numerically using an extended version of the program from
\cite{afmmamp} which will soon be available at \cite{url}.

In the WW approximation, one can use evolution of the twist-2 GPDs to
evaluate Eq.~(\ref{deftw-3}) at a scale different than the initial
scale. This will allow one to study twist-3 effects for the first time
with a varying scale $Q^2$. Since the twist-3 NLO coefficient
functions are unknown (see \cite{kivmach} for a recent calculation of
the non-singlet quark sector), I will restrict myself to a LO analysis
in the twist-3 sector.

Since twist-3 is entirely expressible through the twist-2 in the WW
approximation, I will use the same $t$ dependence for the twist-3 DVCS
amplitudes as I used in the twist-2 case. Having completed the
specifications of the twist-2 and twist-3 sectors, I can now move on
and discuss twist-3 effects in DVCS observables which will be done in
the next section.

\section{DVCS observables: Twist-2 vs. Twist-3 results}
\label{resultsob}

In the following, I will discuss twist-3 effects in the SSA for four
experimental settings: HERA, EIC, HERMES and CLAS and the CA for HERA,
EIC and HERMES. Note that for illustrative purposes I will not only
discuss twist-3 effects in LO but include the LO twist-3 amplitudes
together with the NLO twist-2 and NLO kinematic power corrections
terms. This is actually not legitimate since one is mixing different
orders of $\alpha_s$.  However, it serves both to illustrate the LO vs. NLO
effects without genuine twist-3 effects and to set an upper limit on
the twist-3 corrections in NLO, since the twist-3 NLO coefficient
functions will not induce larger corrections than NLO in the twist-2
sector. This can be seen from the LO twist-3 coefficient function
Eq.~(\ref{coefftw-3}) which has only a regulated logarithmic
singularity instead of a regulated simple pole as the LO twist-2
coefficient function (see for example \cite{bemu4,afmmamp}). Hence it
can be expected that the singularity structure of the NLO twist-3 will
also be less severe than in the twist-2 coefficient function and the
LO twist-3 effects will give a reliable upper bound for the NLO case.

\subsection{HERA}
\label{hera}

In this section, I discuss the effects of LO twist-3 effects on the CA
and SSA in HERA kinematics with $\sqrt{s} = 319~\mbox{GeV}$ i.e.
$27.6~\mbox{GeV}$ unpolarized/polarized positrons/electrons and
$920~\mbox{GeV}$ unpolarized protons.  Since it will be difficult for
either ZEUS or H1 to measure a $t$ distribution for DVCS, I will only
discuss the CA and SSA integrated over $t$. Since the largest $t$ for
which the HERA experiments still have a signal is about
$\simeq-1~\mbox{GeV}^2$, I choose a very conservative cut-off in
$t$ of $-0.5~\mbox{GeV}^2$. I have checked that changing the cut-off
to $\simeq-1~\mbox{GeV}^2$ only alters the absolute answers on the order of
$10\%$ as well as leaving the relative twist-3 effect unchanged. Since
I do neither know the acceptance curve in $t$ for the H1 and ZEUS
detector, which induces an additional uncertainty in the answer, nor
for any other of the experiments, for that matter, the chosen cut-off
in $t$ seems to be a sensible choice.

\begin{figure}  
\centering
\mbox{\epsfig{file=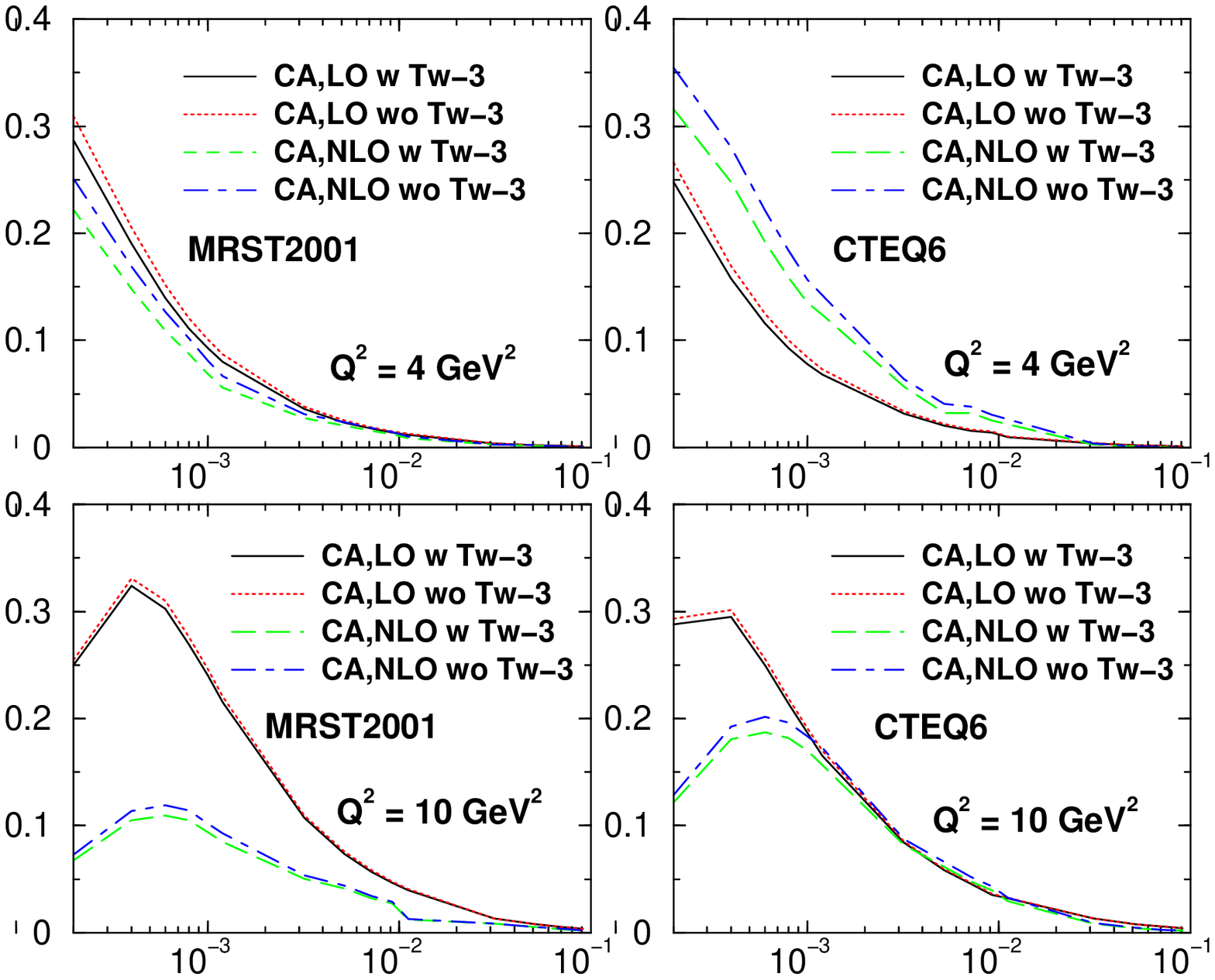,width=8.5cm,height=10.5cm}}
\caption{$t$ integrated CA in HERA kinematics vs. $\Bx$ for two typical values of $Q^2$ and $t_{max}= -0.5~\mbox{GeV}^2$. W stands for ``with'' and WO stands for ``without''.}
\label{heracaqvsx}
\vskip+0.2in
\mbox{\epsfig{file=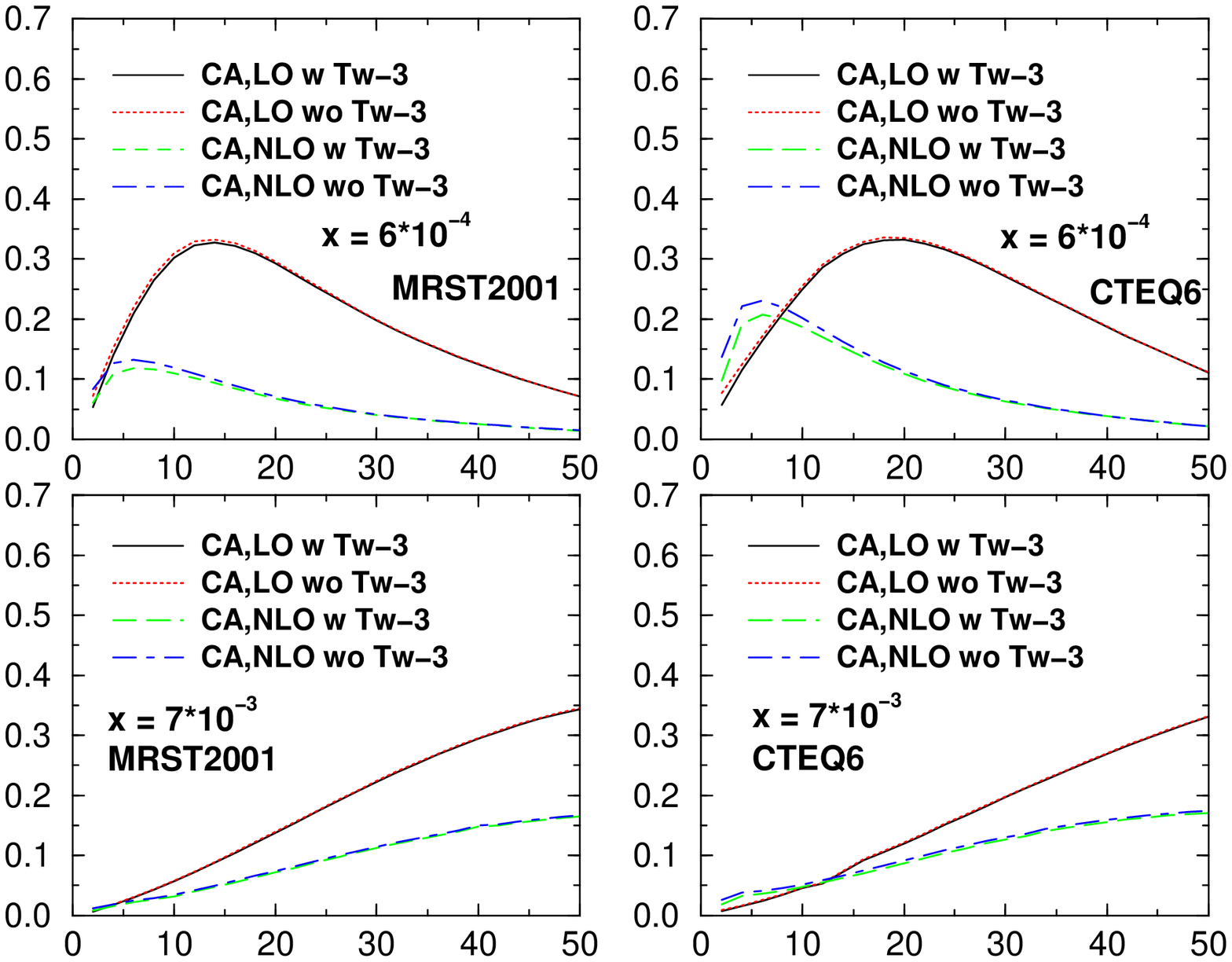,width=8.5cm,height=10.5cm}} 
\caption{$t$ integrated CA in HERA kinematics vs. $Q^2$ for two  typical values of $\Bx$ and $t_{max}= -0.5~\mbox{GeV}^2$.}
\label{heracaxvsq}
\end{figure} 

As can directly be seen from the Figs.~\ref{heracaqvsx} and
\ref{heracaxvsq}, the twist-3 effect in the CA are entirely negligible
at HERA for both the MRST2001 and CTEQ6L LO parameterizations.
Furthermore, the two distributions give the same answer within about
$10\%$. However, when comparing the respective NLO curves one finds
differences of up to $100\%$. They start to disappear for $\Bx\to0.1$ in
the given $Q^2$ range. For small $\Bx$, however, this difference
disappears only for $Q^2>40~\mbox{GeV}^2$. This feature was already
noted earlier \cite{afmmlong} in a pure twist-2 analysis and shown to
be attributable to the very different NLO gluon distributions at
$Q_0$. Also note that the exclusion of kinematic power corrections in
\cite{afmmlong} lead to negative numbers for the CA in HERA kinematics
in contrast to our findings here. This illustrates the importance of
kinematic power corrections for the CA.  The NLO corrections, in
particular at the smallest $\Bx$, are very large and only reduce for
the largest $Q^2$ to about $100\%$.  Again the same was found in a pure
twist-2 NLO analysis \cite{afmmlong} and attributed to a large NLO
gluon contribution in the real part of DVCS amplitudes. However, for
$\Bx>10^{-3}$ and $Q^2\geq10~\mbox{GeV}^2$ the NLO corrections for CTEQ6M
seem to be much smaller than in the case of MRST2001. I will come back
to this point when I discuss the SSA. One word has to be said about
the influence of the D-term on the CA at this point. Based on the
findings about the DVCS amplitudes in \cite{afmmamp} and by explicit
comparison of results with and without a D-term, I conclude that the
influence of the D-term is totally negligible for HERA.

\begin{figure}  
\centering
\mbox{\epsfig{file=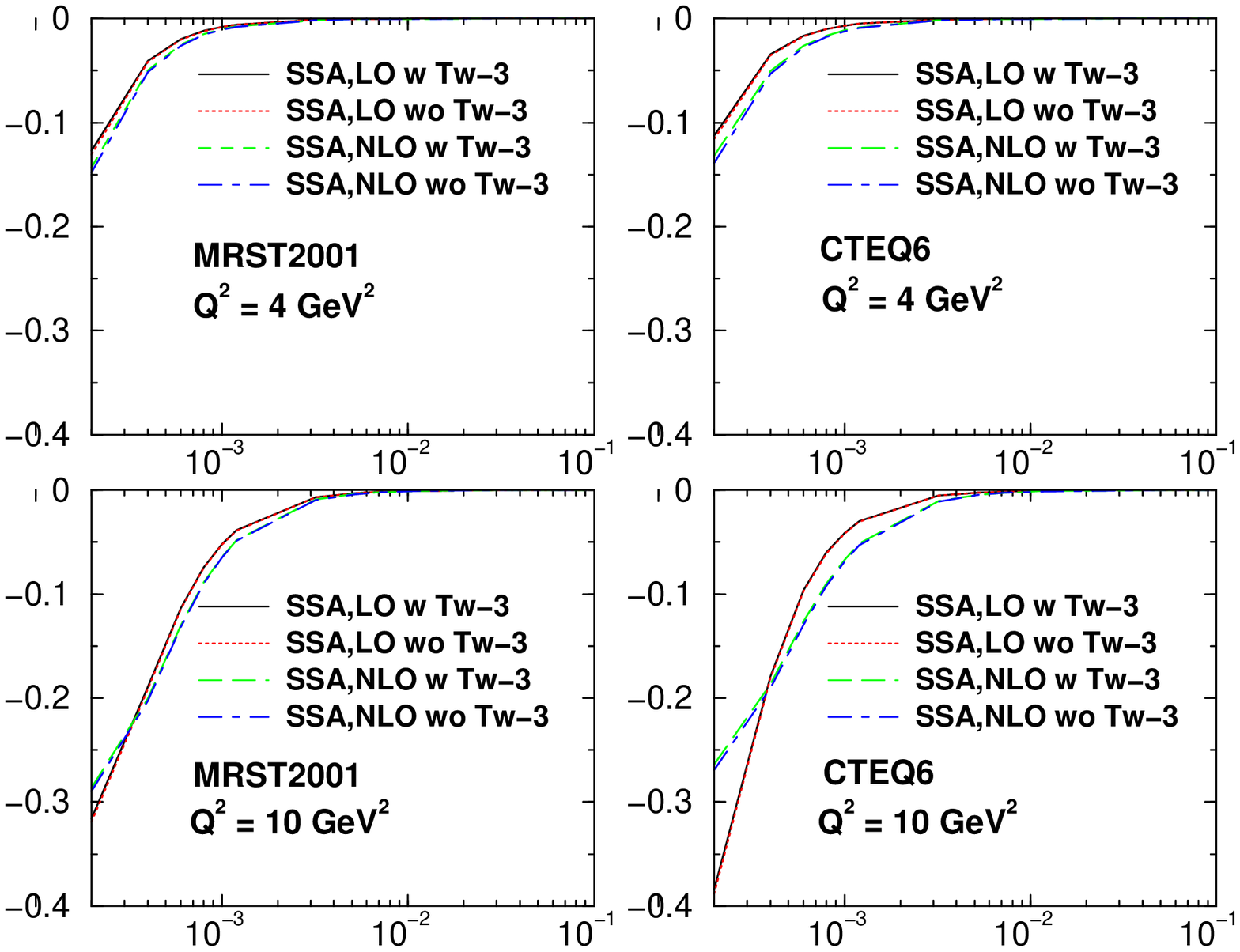,width=8.5cm,height=10.5cm}} 
\caption{$t$ integrated SSA in HERA kinematics vs. $\Bx$ for two typical values of $Q^2$ and $t_{max}= -0.5~\mbox{GeV}^2$. W stands for ``with'' and WO stands for ``without''.}
\label{herassaqvsx}
\vskip+0.2in
\mbox{\epsfig{file=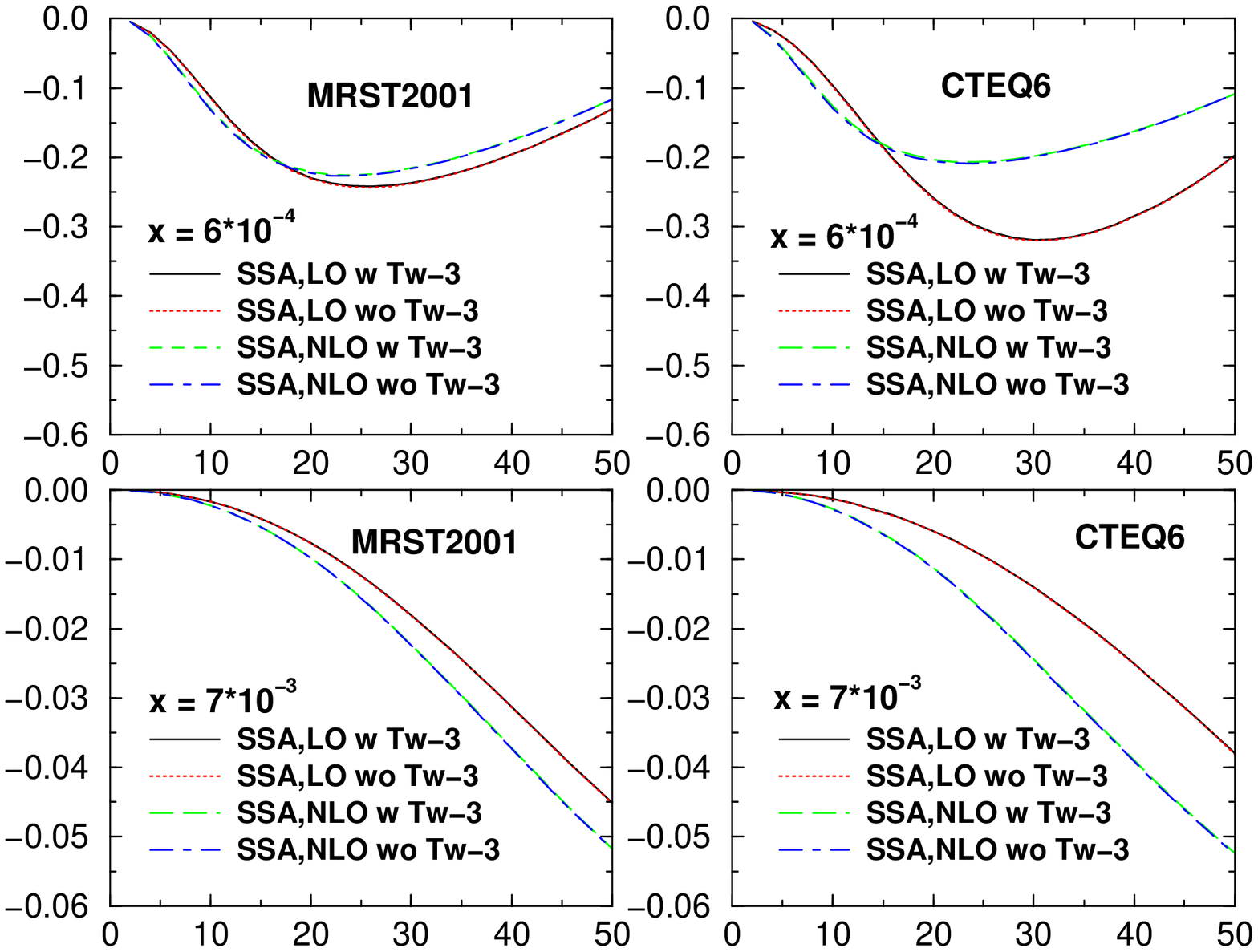,width=8.5cm,height=10.5cm}} 
\caption{$t$ integrated SSA in HERA kinematics vs. $Q^2$ for two  typical values of $\Bx$ and $t_{max}= -0.5~\mbox{GeV}^2$.}
\label{herassaxvsq}
\end{figure}

Turning now to the SSA, that we can see from Figs.~\ref{herassaqvsx}
and \ref{herassaxvsq} that the twist-3 effects are even smaller than
in the case of the CA. We also see that the room for twist-3 effects
in NLO is further reduced compared to the CA. Note that I discuss a
positron rather than an electron beam and therefore the sign of the
asymmetry is negative.

Furthermore, we see that the NLO corrections are typically of the
order of $10-15\%$ but at most $50\%$. This is in agreement with the
results found in \cite{afmmlong} in a pure twist-2 analysis
demonstrating that in contrast to the CA, the SSA is quite insensitive
to kinematic power corrections. We also see that in NLO both MRST2001
and CTEQ6M give almost identical results, however, differ in LO for
large $Q^2$ and $\Bx<10^{-3}$ which simply illustrates the fact that
the LO gluon is larger for CTEQ6L than for MRST2001. In LO, this
difference can only manifests itself after a longer evolution path
since the DVCS amplitude contains only quarks at leading order. In
NLO, where the gluon enters directly on the amplitude level,
differences in the gluon manifest themselves at lower $Q^2$ in
quantities very sensitive to the gluon contribution, like the CA due
to its proportionality to the real part. This is well represented when
comparing the NLO results for CTEQ6M and MRST2001 in the CA and the SSA
at very small $\Bx$.

The results for both LO and NLO suggest that both the CA and SSA
should be easily measurable with fairly high precision at both the H1
and ZEUS experiment!

\subsection{EIC}
\label{eic}

In its current design the electron-ion-collider (EIC) will collide
$1-10~\mbox{GeV}$ electrons from a linear accelerator with
$100-250~\mbox{GeV}$ unpolarized/polarized protons and unpolarized
ions of up to $100~\mbox{GeV}$. Note that the projected luminosity for
one year at the EIC will be larger than for the entire HERA run,
enabling high precision studies of DVCS. For the figures below I chose
a center-of-mass energy of $\sqrt{s} = 63.25~\mbox{GeV}$ which
corresponds to a $5~\mbox{GeV}$ electron beam and a $200~\mbox{GeV}$
proton beam as a sort of average setting for the machine. The $\Bx$
range will thus be between roughly $10^{-3}-10^{-1}$. Naturally, the
higher $\sqrt{s}$ the closer the kinematics will be to HERA and, thus,
also the results.  Since we want to investigate the $\Bx$ region
between HERA and HERMES with an overlap to both experiments, we do not
want to go to the highest energies save for cross-checking HERA
results. Let me start my discussion with the CA once again and then
move on to the SSA. Note that I will apply the same $t$ cuts as in the
case of HERA to be able to compare the two settings. Though the proton
target can be polarized, I will not discuss such an observable since
it was shown in \cite{afmmlong} that such observables are essentially
zero for a collider setting.

\begin{figure}  
\centering
\mbox{\epsfig{file=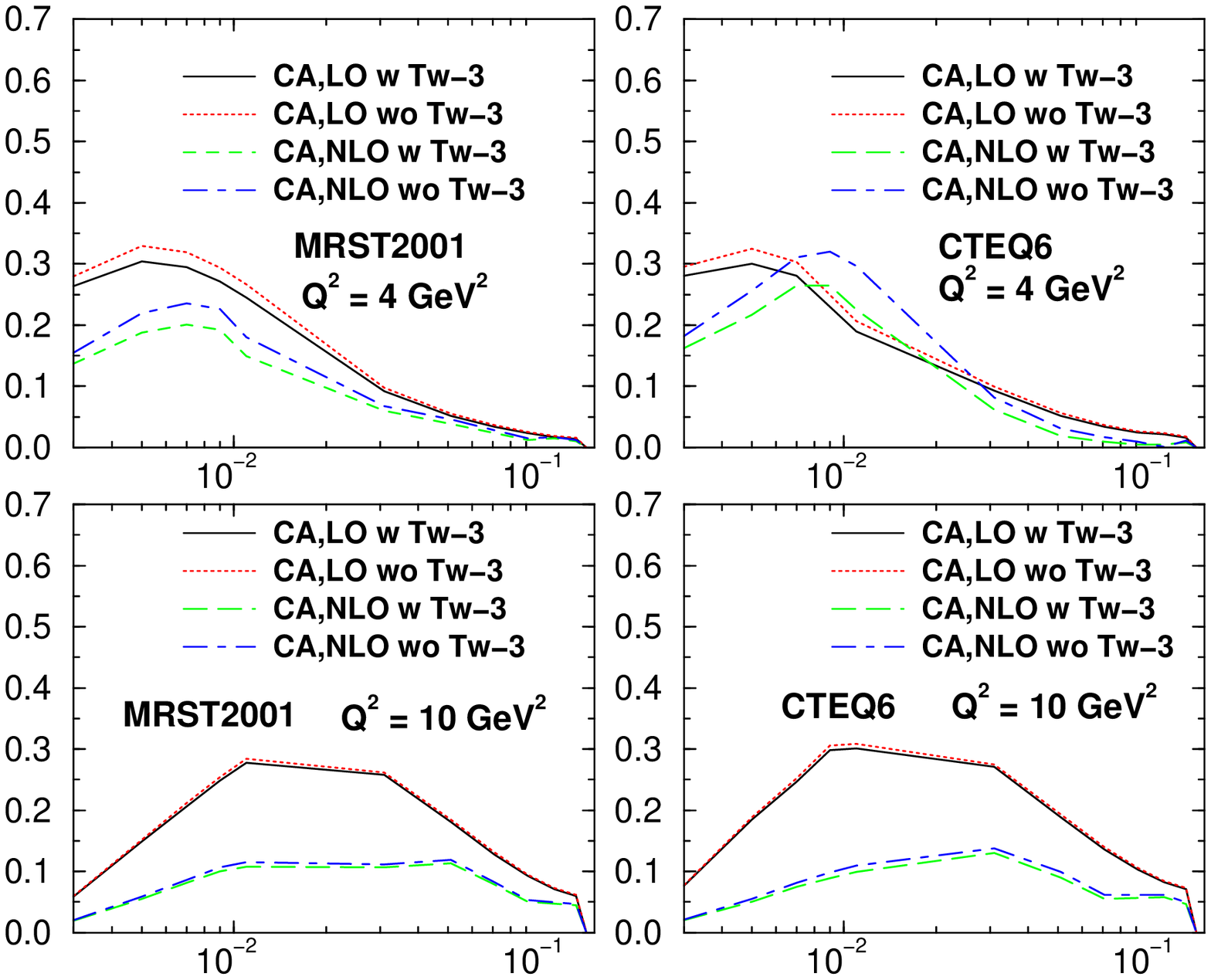,width=8.5cm,height=10.5cm}} 
\caption{$t$ integrated CA in EIC kinematics vs. $\Bx$ for two typical values of $Q^2$ and $t_{max}= -0.5~\mbox{GeV}^2$. W stands for ``with'' and WO stands for ``without''.}
\label{eiccaqvsx}
\vskip+0.2in
\mbox{\epsfig{file=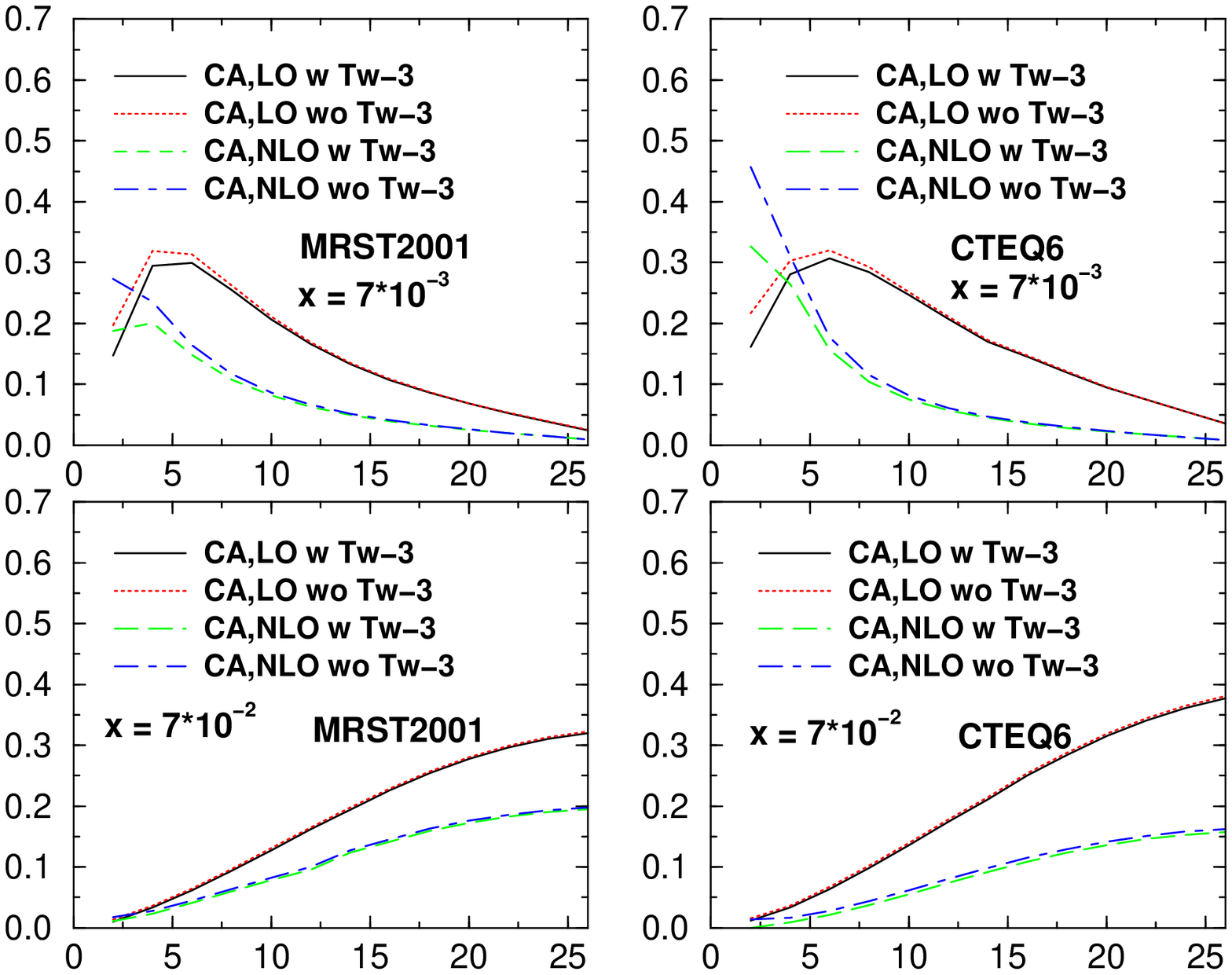,width=8.5cm,height=10.5cm}} 
\caption{$t$ integrated CA in EIC kinematics vs. $Q^2$ for two  typical values of $\Bx$ and $t_{max}= -0.5~\mbox{GeV}^2$.}
\label{eiccaxvsq}
\end{figure}

As can be seen from Figs.~\ref{eiccaqvsx} and \ref{eiccaxvsq}, the
twist-3 effects in LO for the CA are at most $10\%$, except at the
smallest $\Bx$ and lowest $Q^2=2~\mbox{GeV}^2$ where they can reach
around $20\%$ and have to be taken into account when trying to extract
twist-2 GPDs from the data. In NLO the corrections seem larger but
remember that this is only an upper estimate of the actual twist-3
effects in NLO and thus they are not more than $35\%$ at the lowest
$\Bx$ and $Q^2$. It is more likely, however, that they will be of the
same size as the LO result or even smaller.  Note also that the
twist-3 effects quickly vanish for larger $\Bx$ within the entire
$Q^2$ interval, which is mainly a kinematical effect rather than a
dynamical one. The two distributions, CTEQ6 and MRST2001, give very
similar numbers in LO and at NLO for larger $Q^2$, however, differ
strongly at low $Q^2$ as already seen for the HERA setting.  The
relative NLO corrections are again large and follow the same pattern
for both sets as at HERA.  The influence of the D-term on the CA is
again negligible.

\begin{figure}  
\centering
\mbox{\epsfig{file=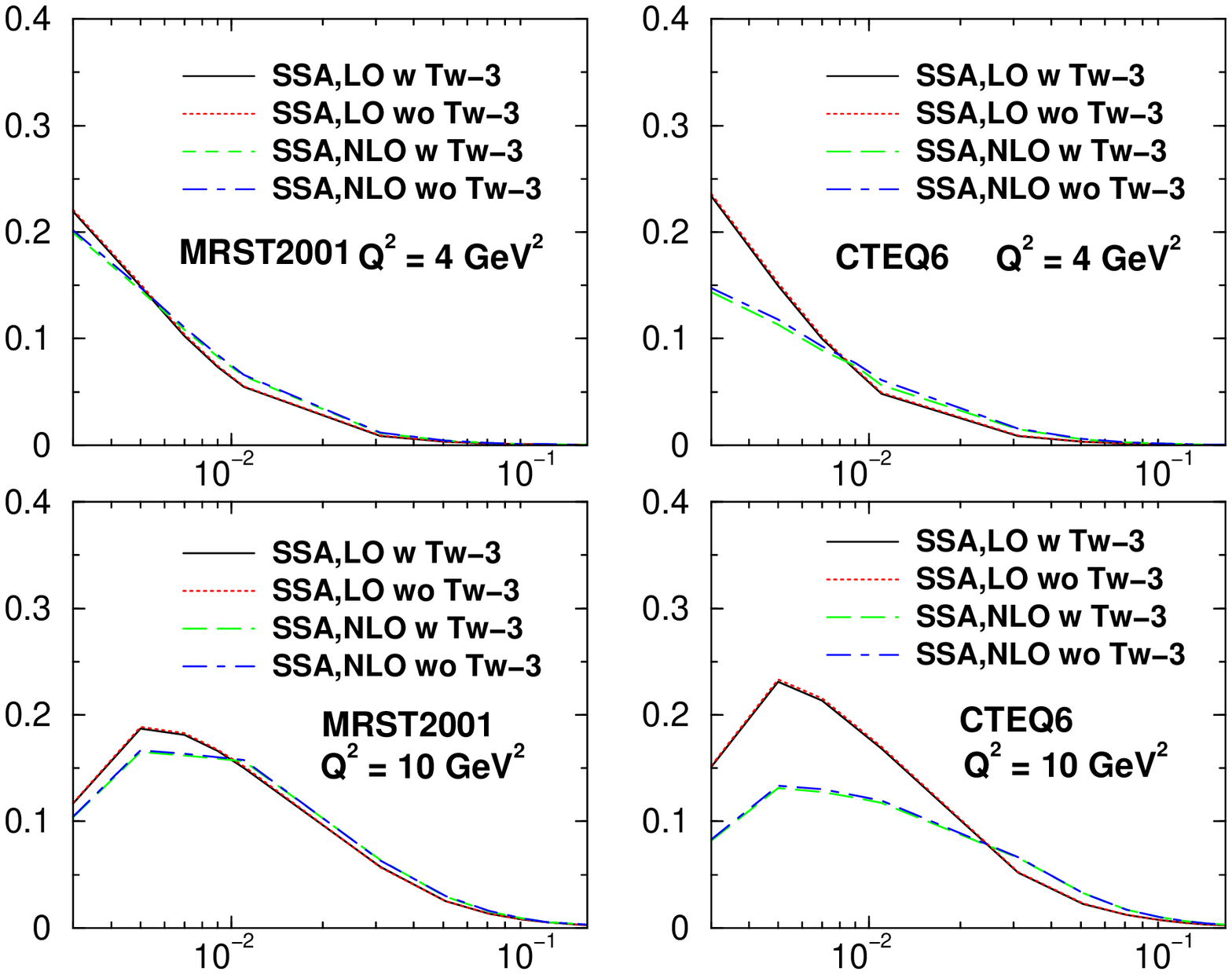,width=8.5cm,height=10.5cm}} 
\caption{$t$ integrated SSA in EIC kinematics vs. $\Bx$ for two typical values of $Q^2$ and $t_{max}= -0.5~\mbox{GeV}^2$. W stands for ``with'' and WO stands for ``without''.}
\label{eicssaqvsx}
\vskip+0.2in
\mbox{\epsfig{file=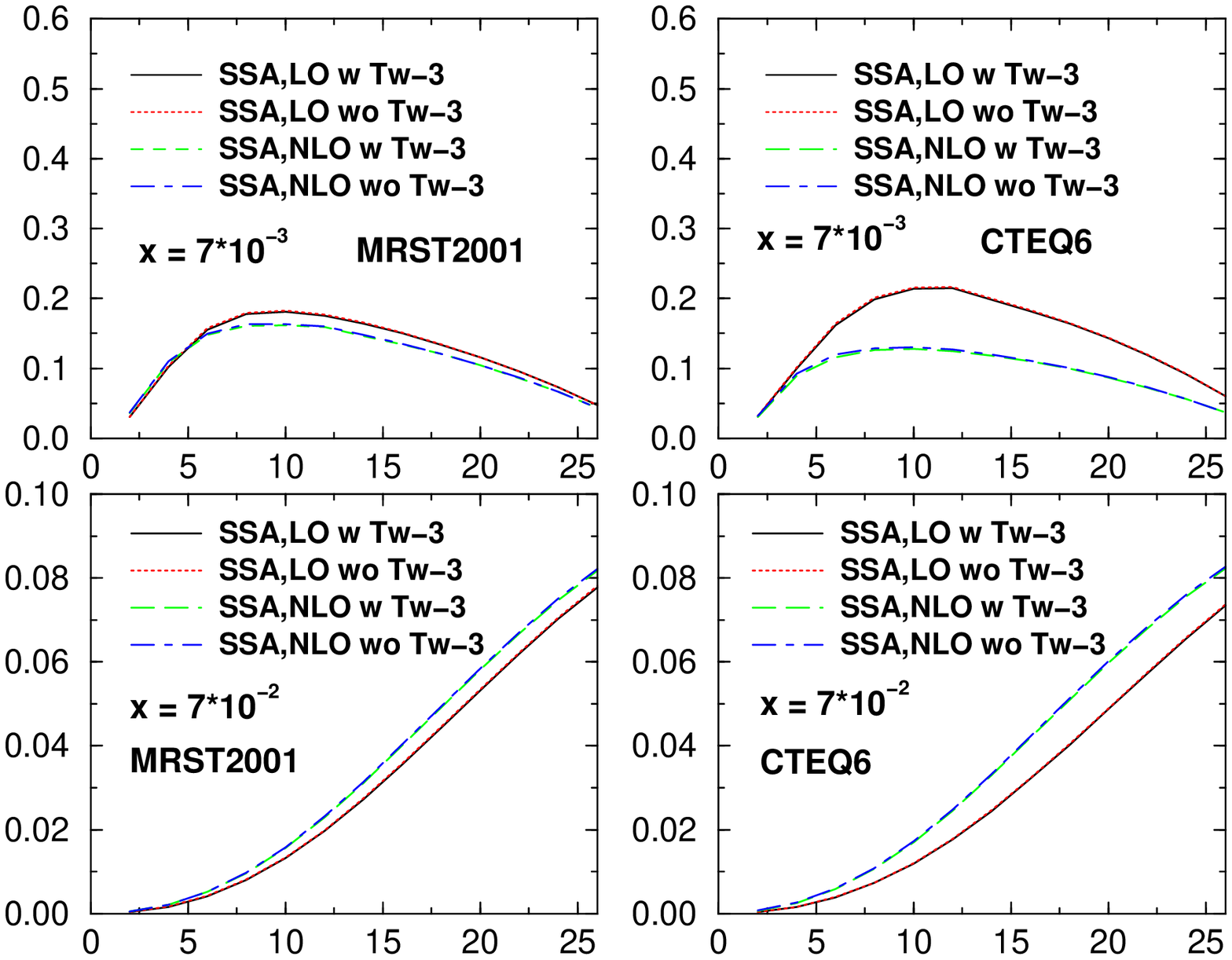,width=8.5cm,height=10.5cm}} 
\caption{$t$ integrated SSA in EIC kinematics vs. $Q^2$ for two  typical values of $\Bx$ and $t_{max}= -0.5~\mbox{GeV}^2$.}
\label{eicssaxvsq}
\end{figure}

When looking at the SSA in Figs.~\ref{eicssaqvsx} and
\ref{eicssaxvsq}, one notices that the twist-3 effects are basically
zero as in the case of HERA and that the NLO corrections are very
moderate and of the same size for both sets as in the HERA case. Hence
they can be safely neglected in a GPD extraction. Note that the shape
of the SSA in $\Bx$ and $Q^2$ is the mirror of the one at HERA since
the EIC uses an electron rather than a positron beam.

In conclusion one can say that except at low $Q^2$ and the smallest
$\Bx$ in the CA or a similar asymmetry, the twist-3 effects can be
safely neglected and that the size is basically the same as in the
case of HERA. These are very encouraging signs that, together with the
high luminosity, DVCS will be measured with high precision at the EIC.
Therefore, we will be able to {\it reliably} extract the leading
twist-2 GPD $H$ with high precision in a very broad range of $\Bx$ and
$Q^2$ from the EIC DVCS data!

\subsection{HERMES}
\label{hermes}

In the following I will discuss the fixed-target experiment HERMES
with a center-of-mass energy of $\sqrt{s} = 7.2~\mbox{GeV}$.
This allows, broadly speaking, to access a region in $\Bx$ of about
$0.05 - 0.3$ with $Q^2$ from $1-9~\mbox{GeV}^2$. HERMES uses the
electron/positron beam from HERA with an energy of $E =
27.6~\mbox{GeV}$. The gas target can be either unpolarized/polarized
protons or unpolarized nuclei. I will not discuss observables with a
polarized target for HERMES, since there is no clear leading DVCS
amplitude, such as $H$ in the case of the CA and SSA, and thus the
disentangling of the various contributing GPDs is supremely difficult.
In order to allow a comparison with the collider experiments, I once
more choose a $t_{max}$ of $-0.5~\mbox{GeV}^2$ which is also not an
unrealistic choice given the fact that the average $t$ for HERMES is
about $-0.25~\mbox{GeV}^2$.

\begin{figure}  
\centering
\mbox{\epsfig{file=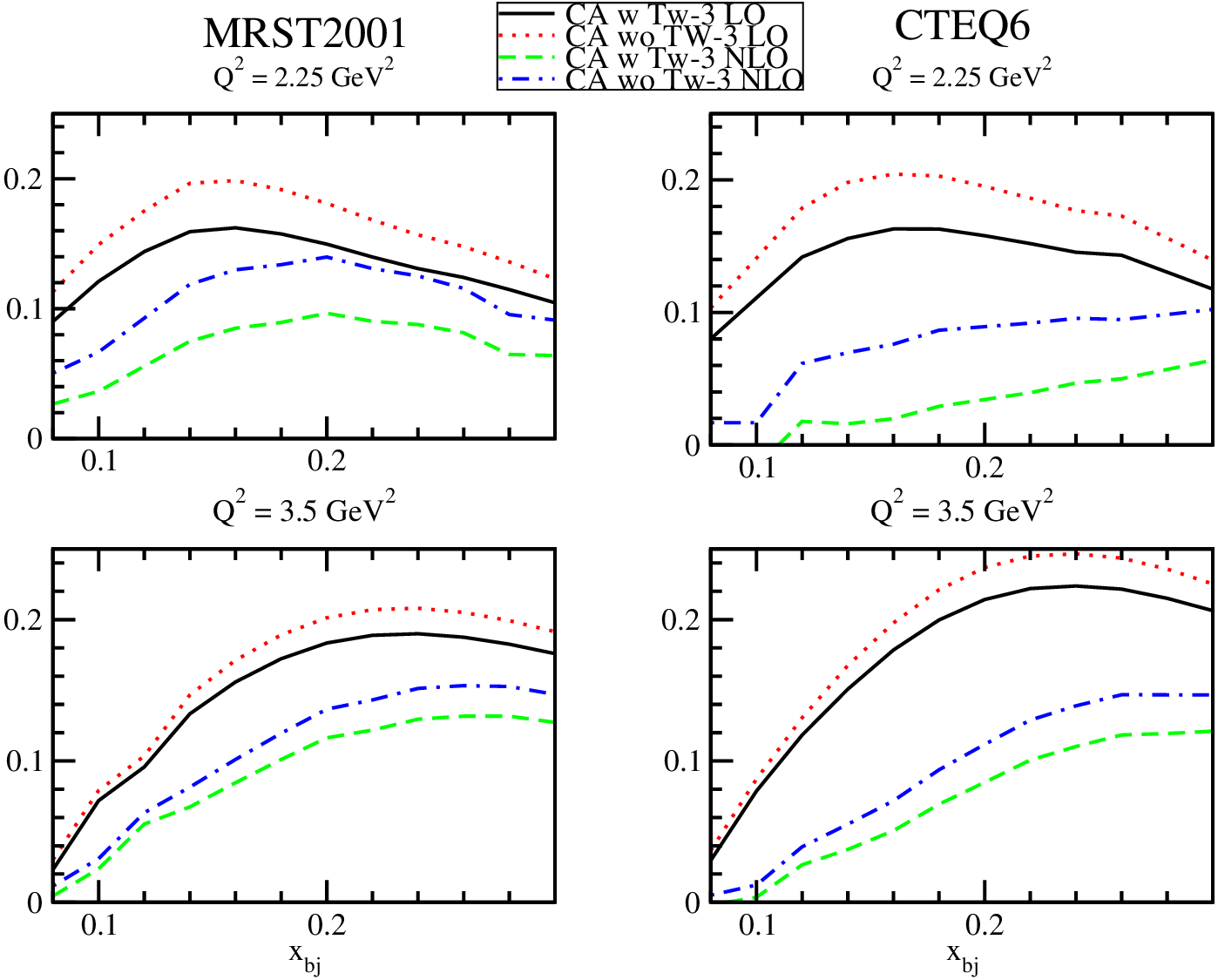,width=8.5cm,height=10.5cm}} 
\caption{$t$ integrated CA in HERMES kinematics vs. $\Bx$ for two typical values of $Q^2$ and $t_{max}= -0.5~\mbox{GeV}^2$. W stands for ``with'' and WO stands for ``without''.}
\label{hermescaqvsx}

\vskip+0.2in
\mbox{\epsfig{file=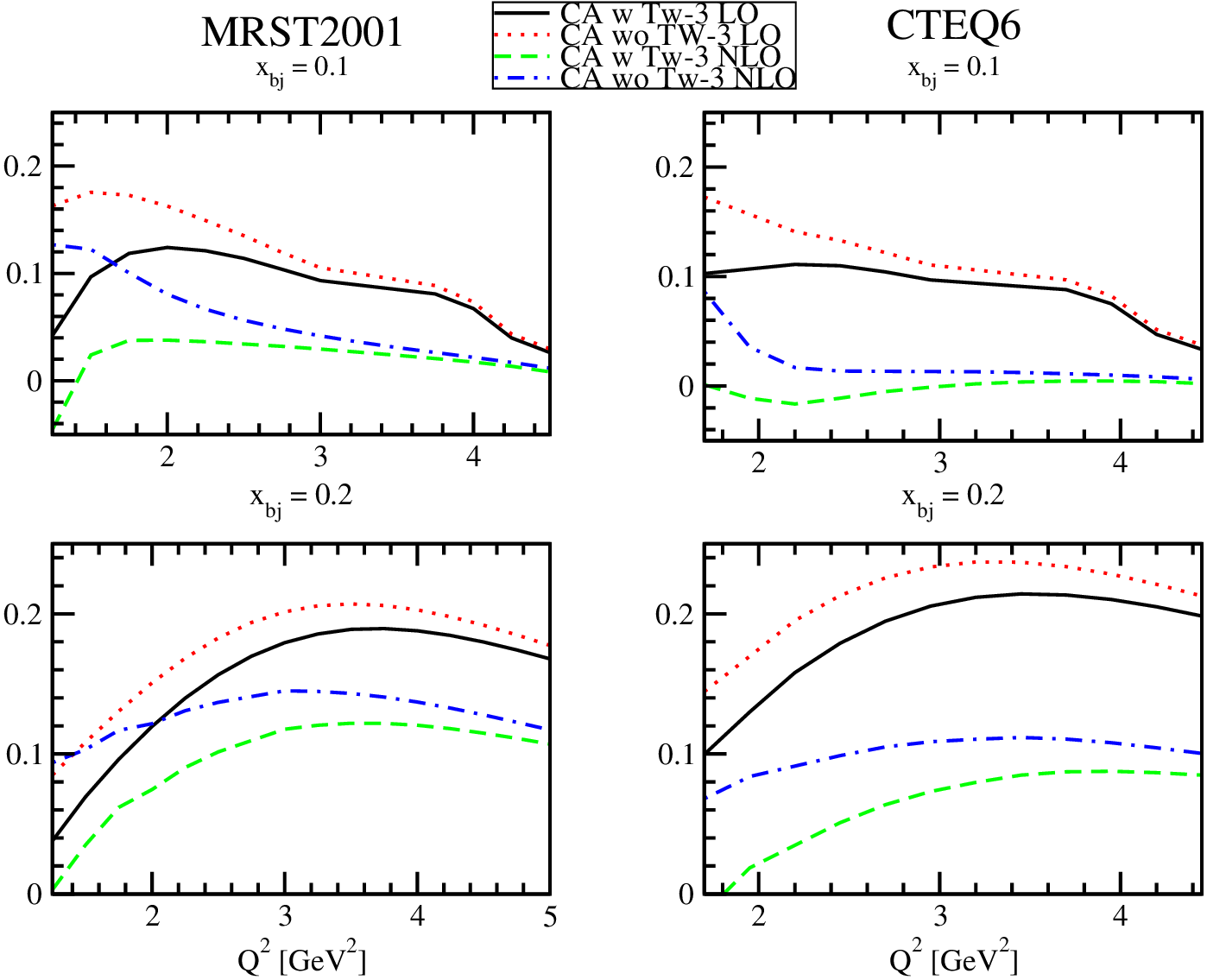,width=8.5cm,height=10.5cm}} 
\caption{$t$ integrated CA in HERMES kinematics vs. $Q^2$ for two  typical values of $\Bx$ and $t_{max}= -0.5~\mbox{GeV}^2$.}
\label{hermescaxvsq}
\end{figure}

The CA, as can be seen from Figs.~\ref{hermescaqvsx} and
\ref{hermescaxvsq}, receives larger twist-3 corrections in LO than the
CA at HERA or EIC. However, except, at the lowest values of $Q^2$ and
smallest $\Bx$ where they can be as large as factor of $4$, the
corrections are generally speaking $15\%$ or less. Note that as $Q^2$
increases the twist-3 corrections rapidly disappear in both LO and
NLO.  The LO results between the two sets agree very nicely but there
is quite a difference in NLO. The NLO corrections themselves are again
quite large but not larger than at HERA or EIC. In fact, for larger
$Q^2$ and larger $\Bx$ they are quite small. Note that when averaging
the LO and NLO results with kinematic power corrections for the CA for
both sets over $Q^2$ and $\Bx$ one obtains the same numbers as in
\cite{afmmms} while the number for LO with full twist-3 is about $0.1$
neatly interpolating between the LO and NLO result of $0.12$ and
$0.09$ respectively. This compares very favorably with the
experimental HERMES result of $0.11\pm0.04\pm0.04$ \cite{hermnew} for $\langle x
\rangle = 0.12, \langle Q^2 \rangle = 2.8~\mbox{GeV}^2, \langle t \rangle = -0.27~\mbox{GeV}^2$.
One can see that the averaging process washes out any differences
between the two GPD sets, which, however, were not that tremendous to
begin with. A word about the D-term and its influence on the CA is in
order at this point.  Recently \cite{gpdlat1,gpdlat2}, the first
lattice results on the coefficient of the D-term were obtained and
found to differ from the prediction of the chiral-quark-soliton model
\cite{vanderhagen1} quite substantially. The respective calculations
were done at different normalization points ($\mu =2 ~\mbox{GeV}$ for
the lattice i.e. within HERMES kinematics, and $\mu =0.6 ~\mbox{GeV}$
for the chiral-quark-soliton model). Evolution itself cannot account
for the observed difference of about a factor of $4$.  When studying
the LO and NLO evolution of the D-term using the GRV98 PDF
\cite{grv98} with the above GPD Ansatz one finds that in both LO and
NLO the quark D-term is reduced by about $30\%$ from the respective
input scale of $Q_0=0.51~\mbox{GeV}$ (LO) and $Q_0=0.63~\mbox{GeV}$
(NLO) to $Q=2~\mbox{GeV}$ (the difference between LO and NLO is about
$2\%$), leaving still a factor of about $3$ between the two results
modulo the uncertainty associated with a gluonic D-term which migh,
given the right sign and size, be able to account for the observed
difference.  GRV98 was used since the input scales are very close to
the one of the chiral-quark-soliton model. When studying the
importance of the D-term for DVCS observables I find that if the
D-term were either omitted or its size reduced by a factor of about
$3-4$, the CA would become so small that it would no longer be in good
agreement with the data.  However, using a different Ansatz for
twist-2 GPDs based on a double distribution model (see for example
\cite{rad1}) and neglecting evolution effects, the authors of
\cite{bemu4} describe the CA without a D-term. It was shown in
\cite{afmmms}, though, that this type of double distribution Ansatz as
chosen in \cite{bemu4}, cannot describe the DVCS data in either LO or
NLO when evolution effects are taken into account! The situation will
unfortunately remain unresolved until better fixed target data will
become available.

\begin{figure}  
\centering
\mbox{\epsfig{file=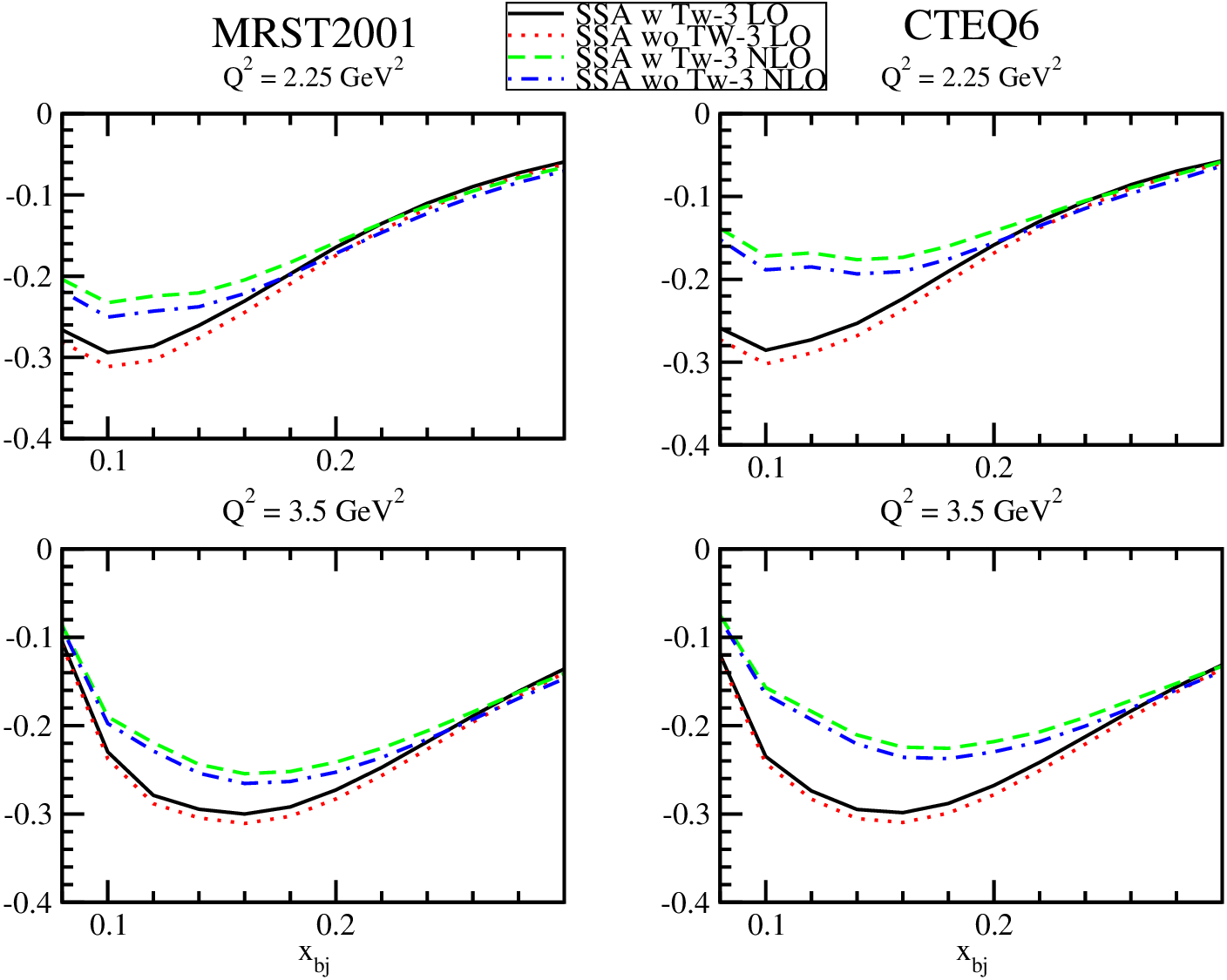,width=8.5cm,height=10.5cm}} 
\caption{$t$ integrated SSA in HERMES kinematics vs. $\Bx$ for two typical values of $Q^2$ and $t_{max}= -0.5~\mbox{GeV}^2$. W stands for ``with'' and WO stands for ``without''.}
\label{hermesssaqvsx}
\vskip+0.2in
\mbox{\epsfig{file=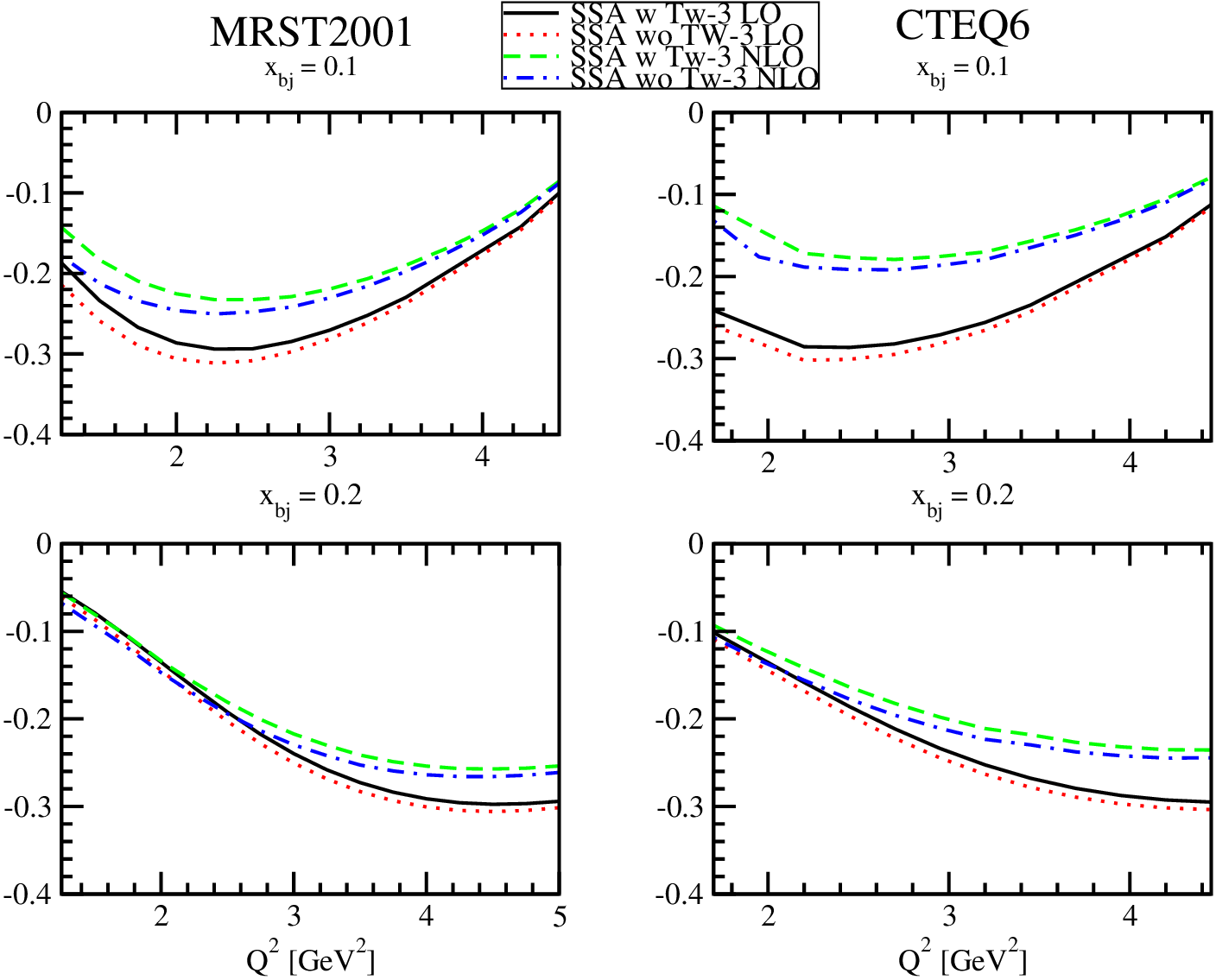,width=8.5cm,height=10.5cm}} 
\caption{$t$ integrated SSA in HERMES kinematics vs. $Q^2$ for two  typical values of $\Bx$ and $t_{max}= -0.5~\mbox{GeV}^2$.}
\label{hermesssaxvsq}
\end{figure}

When turning to the SSA in Figs.~\ref{hermesssaqvsx} and
\ref{hermesssaxvsq}, one can see that the twist-3 effects in LO are at
most $10\%$ and that the NLO corrections are, as in the case of HERA
and EIC, very moderate and at most about $35\%$. The results of the two
sets in LO are virtually identical and still within $20\%$ at NLO. When
averaged over $Q^2$ and $\Bx$ the results of the two sets do not
differ any longer and reproduce the LO and NLO results of
\cite{afmmms} $-0.28$ and $-0.23$ as they should since the model is
the same. This again compares favorably with the experimental result
of $-0.21\pm0.04\pm0.04$ \cite{hermssa} for virtually the same average
kinematics as the CA.

In conclusion, one can say that higher twist effects can be neglected
for the SSA at HERMES and thus it can serve as a tool for GPD
extraction. The CA is much more sensitive to twist-3 effects, however,
they are still small enough that they can be neglected given the
accuracy of the data, except for the lowest $Q^2$ and $\Bx$ values.
This implies that for about $Q^2>2-2.5~\mbox{GeV}^2$ the CA can also
be used for a GPD extraction or at the very least as a cross check to
fits from smaller $\Bx$ and the HERMES SSA. The GPD model used in this
study already produces very favorable agreement with the SSA and CA
data without resorting to a fit and can thus serve as a basis for a
successful parameterization.

\subsection{CLAS}
\label{clas}

The CLAS experiment is a fixed target experiment with very high
luminosity but a low center of mass energy. I will first investigate
an electron beam of $E=4.3~\mbox{GeV}$ and then one with
$E=5.75~\mbox{GeV}$ corresponding to the energies at the first and
second CLAS run respectively. Here I will concentrate on the SSA and
omit the CA or a similar asymmetry due to the mentioned difficulties
CLAS has or will have with these type of asymmetries as explained in
Sec.~\ref{dvcsobs}. Also, I will only discuss the set of GPDs
generated from MRST2001 since the $Q_0=1~\mbox{GeV}$ is low enough,
compared to the one from CTEQ6 of $Q_0=1.3~\mbox{GeV}$, to have a
meaningful range in $\Bx$ and $Q^2$ for both CLAS settings.

\begin{figure}  
\centering
\mbox{\epsfig{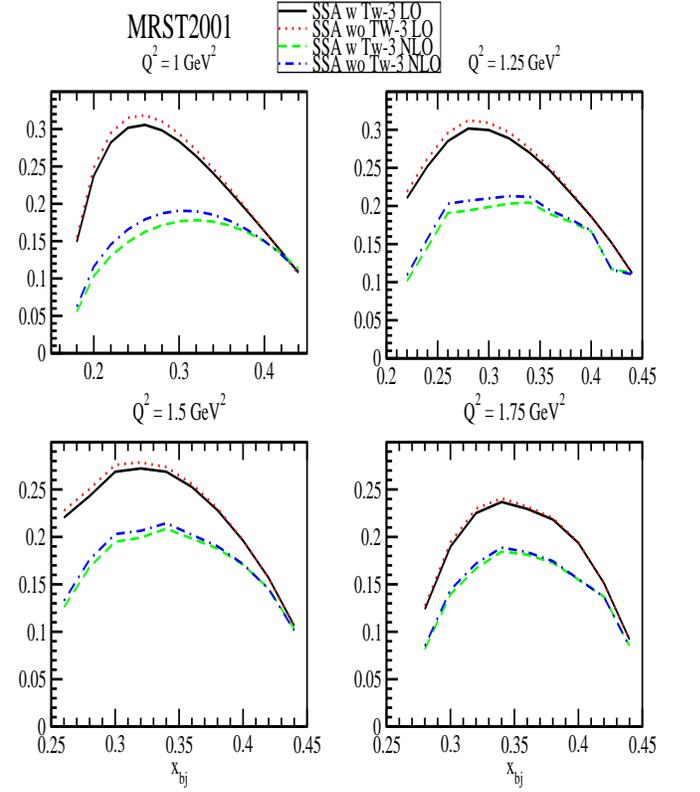}} 
\caption{$t$ integrated SSA in CLAS kinematics vs. $\Bx$ for four typical values of $Q^2$ and $t_{max}= -0.25~\mbox{GeV}^2$. W stands for ``with'' and WO stands for ``without''.}
\label{ssajlablowen}
\end{figure}

Let us start with the lower energy setting, where I have chosen a
$t_{max} = -0.25~\mbox{GeV}$ to get as large a range in $\Bx$ and
$Q^2$ as possible. As one can see from Fig.~\ref{ssajlablowen} the
twist-3 effects in LO are even for such a low energy as CLAS has less
than $10\%$ and thus basically negligible! This fits in nicely with the
measured twist-3 effect at CLAS \cite{clas1} which is about $10\%$ of
the measured SSA for the central value but is compatible with zero
within the experimental errors. Furthermore, the NLO effects are at
most $50\%$ and typically around $20\%$ and thus not as large as one
might have feared for such low $Q^2$ values. This is mainly due to the
fact that the influence of the gluon on the amplitude in NLO at large
$\Bx$ is not as pronounced as at smaller $\Bx$ where its importance
grows quickly. Notwithstanding this fact, the usage of perturbation
theory at such small $Q^2$ remains still questionable on general
grounds.  However, one can definitely say that the twist-2 handbag
contribution to DVCS is the leading contribution to the SSA at CLAS.
This alone is quite an amazing statement given that one would
naively have expected that at these energies higher twist
contributions would be the dominant ones. When averaging over $\Bx$
and $Q^2$ one obtains a value for the SSA in average CLAS kinematics
($\langle x \rangle = 0.19, \langle Q^2 \rangle = 1.31~\mbox{GeV}^2, \langle t \rangle =
-0.19~\mbox{GeV}^2$) of about $0.2$ in LO and about $0.14$ in NLO
which is, at least in LO, in good agreement with the experimental
value of $0.202\pm0.021\pm0.02$ \cite{clas1}.

\begin{figure}  
\centering
\mbox{\epsfig{file=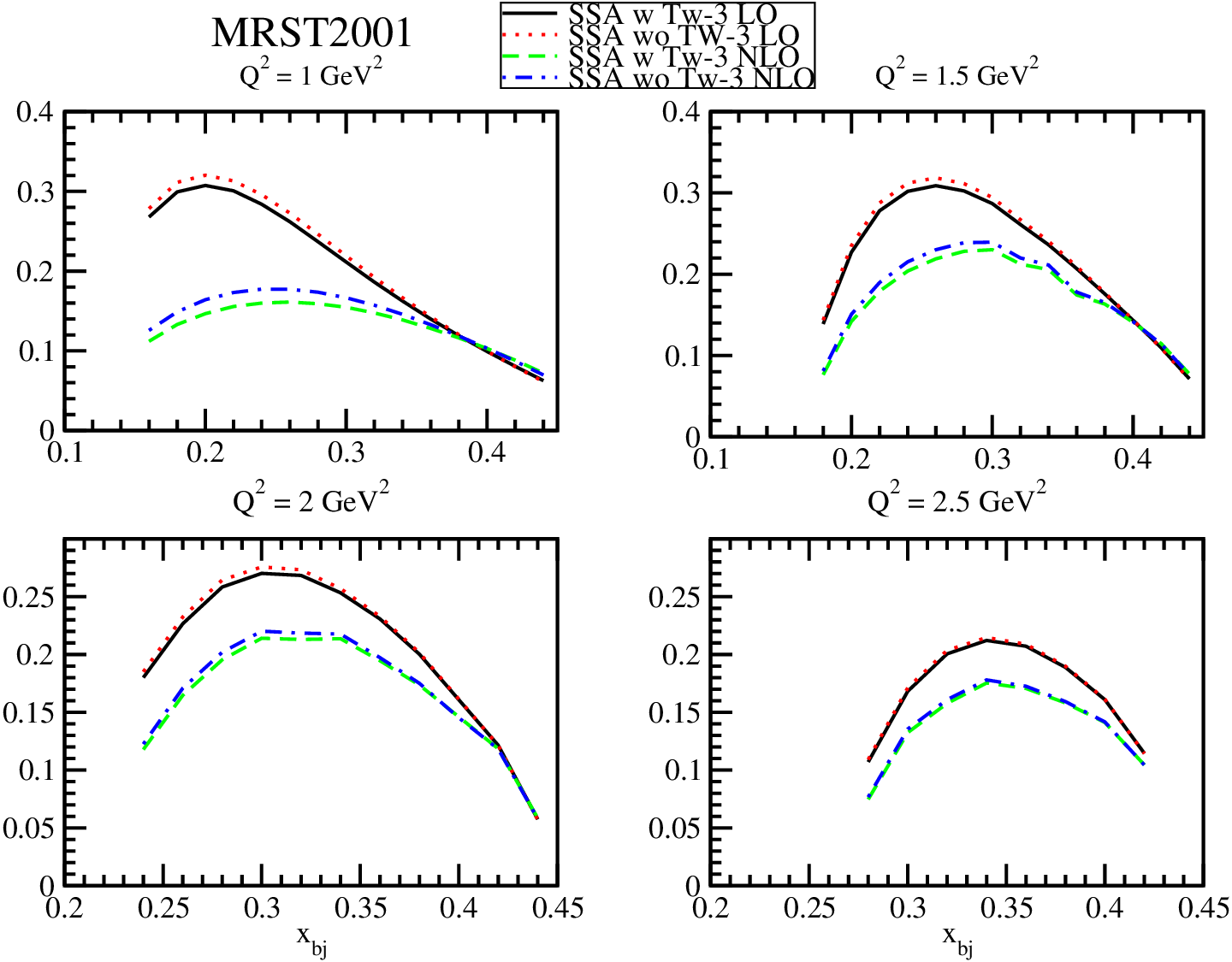,width=8.5cm,height=10.5cm}} 
\caption{$t$ integrated SSA in CLAS kinematics vs. $\Bx$ for four typical values of $Q^2$ and $t_{max}= -0.25~\mbox{GeV}^2$. W stands for ``with'' and WO stands for ``without''.}
\label{classsaqvsxlt}
\vskip+0.2in

\mbox{\epsfig{file=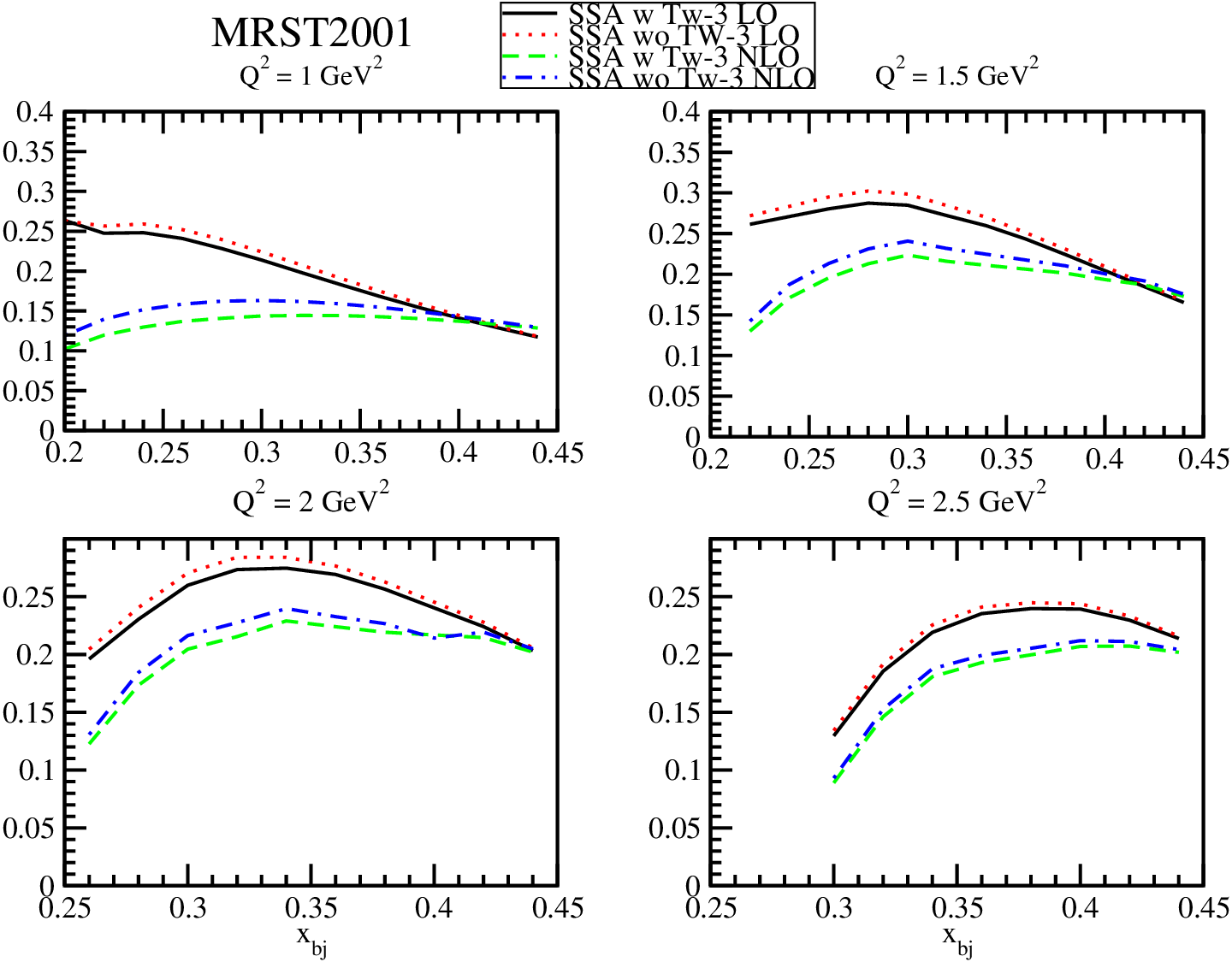,width=8.5cm,height=10.5cm}} 
\caption{$t$ integrated SSA in CLAS kinematics vs. $\Bx$ for four typical values of $Q^2$ and $t_{max}= -0.5~\mbox{GeV}^2$. W stands for ``with'' and WO stands for ``without''.}
\label{classsaqvsx}


\end{figure}

\begin{figure}  
\centering
\mbox{\epsfig{file=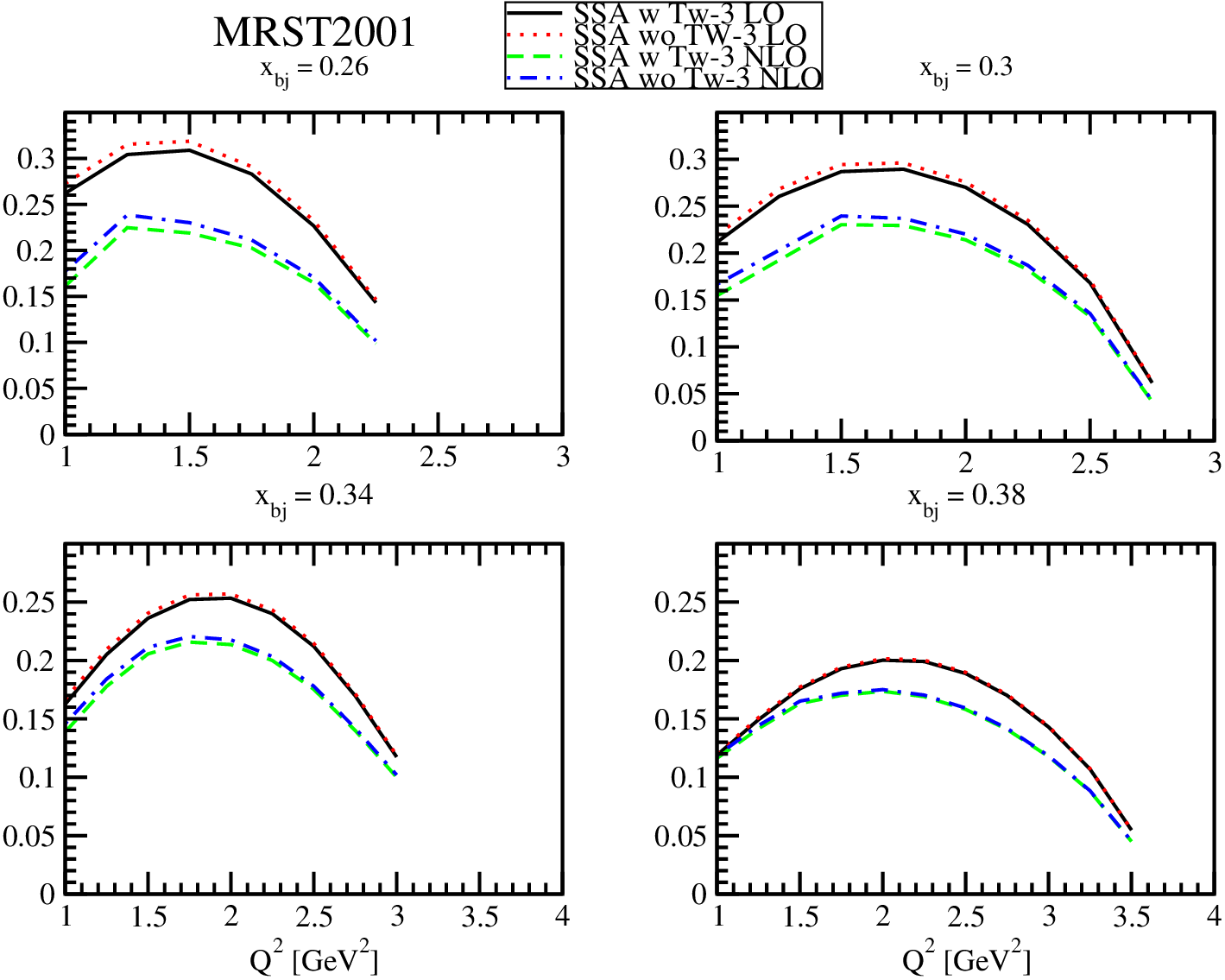,width=8.5cm,height=10.5cm}} 
\caption{$t$ integrated SSA in CLAS kinematics vs. $Q^2$ for four typical values of $\Bx$ and $t_{max}= -0.25~\mbox{GeV}^2$.}
\label{classsaxvsqlt}
\vskip+0.2in

\mbox{\epsfig{file=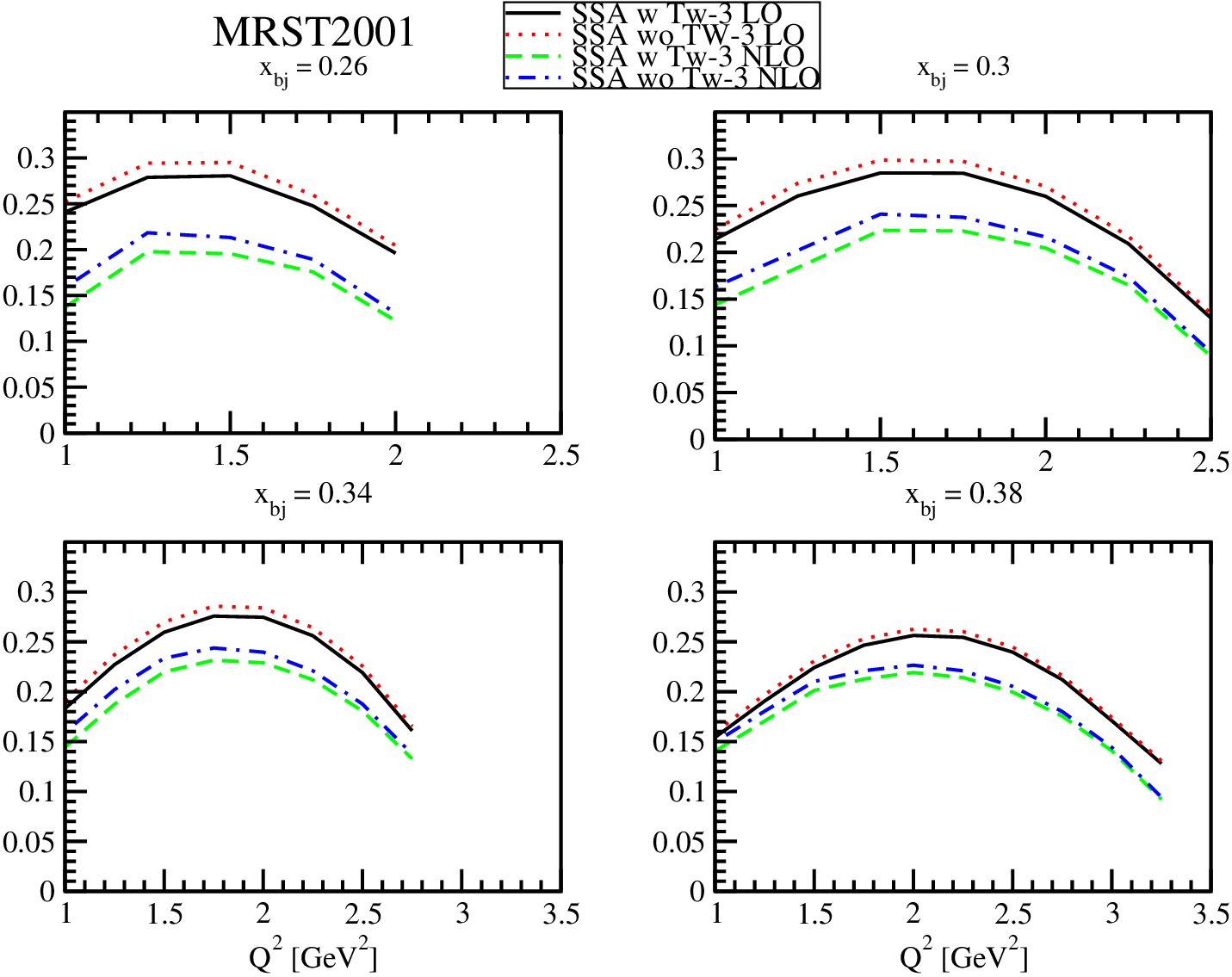,width=8.5cm,height=10.5cm}} 
\caption{$t$ integrated SSA in CLAS kinematics vs. $Q^2$ for four typical values of $\Bx$ and $t_{max}= -0.5~\mbox{GeV}^2$.}
\label{classsaxvsq}
\end{figure}

In the higher energy setting, I have introduced two different
$t_{max}$ values, in order to both compare to the lower energy setting
and demonstrate how the SSA changes for a drastic change in cut-off
for $t$. As can be seen from Figs.~\ref{classsaqvsxlt},
\ref{classsaqvsx}, \ref{classsaxvsqlt} and \ref{classsaxvsq} the
twist-3 contributions are again very small in both LO and NLO for both
cuts in $t$ and are at most $10\%$ which is in agreement with the value
at lower energy. The NLO corrections are mostly moderate except for
the lowest values of $\Bx$ and $Q^2$ as seen at lower energies. The
distribution in $\Bx$ for different $Q^2$ between the lower and higher
energy setting at the same $t_{max}$ (Figs.~\ref{ssajlablowen} and
\ref{classsaqvsxlt}) shows the distributions to be very similar both
in shape and size. When comparing the different cuts in $t$ for the
higher energy setting (Figs.~\ref{classsaqvsxlt} and \ref{classsaqvsx}
and Figs.~\ref{classsaxvsqlt} and \ref{classsaxvsq}) one notices that
the distributions in $\Bx$ with the higher $t$ cut are wider and thus
flatter than the one for a lower cut in $t$. The maxima of the curves
move also to larger values of $\Bx$. The $Q^2$ distributions are
virtually unaltered. There is an overall tendency for the maxima to be
somewhat higher for the higher $t$ cut, but only by at most $10\%$.

We thus see that different cuts in $t$ have only a marginal effect in
the size and distribution of the SSA. When further averaging over
$\Bx$ and $Q^2$ the sensitivity will be further reduced. In fact, as
one can see from the figures, cuts in $\Bx$ and $Q^2$ will have a much
bigger effects than the one in $t$!

Unfortunately, there is no published data yet with which to compare
and without knowing the average kinematics, let alone the experimental
acceptance, it is impossible to make a sensible prediction at this
point.

\section{Phenomenological parameterization of the GPD $H$}
\label{paramet}

Let me now turn to a sensible, phenomenological parameterization of
the $X$, $\zeta$ and $t$ dependence of the leading twist-2 GPD $H$ at a
low normalization point $Q_0$. First, I will give the parameterization
of $H$ in LO and NLO and then justify it based on the available data.

As is clear from the preceding section, the NLO parameterizations seem
to work very well in their current form except for CLAS. However, for
CLAS the GPD $H$ will not be the leading one anymore as it is for
HERA, EIC and HERMES. In fact, the contributions from other GPDs could
be set to zero without changing the HERA and only by a few percent the
HERMES results. Since I only want to make a statement about $H$, I
will restrict myself to a good description of the data from H1, ZEUS
and HERMES.

Based on the analysis carried out in \cite{afmmms}, the MRST2001 NLO
PDF parameterization at $Q_0=1~\mbox{GeV}$ with
$\Lambda^{N_f=4,NLO}_{QCD}=323~\mbox{MeV}$ using the prescriptions of
Eqs.~(\ref{fwd}),(\ref{ajmerbl}) and (\ref{defCs}) does the best job in
describing the DVCS data from H1, ZEUS and HERMES. There is no need to
change the NLO parameterization of Sec.~\ref{gpdmodel}.

The story is different for LO. The LO results using
Eqs.~(\ref{fwd}),(\ref{ajmerbl}) and (\ref{defCs}) are consistently
above the DVCS data save for CLAS. A way to find a LO parameterization
giving a good description of the data is to vary the shift parameter
$a$ in $\zeta$. The shift parameter is given by the number in front of $\zeta$
in the argument of the forward PDF $(X-a\zeta)/(1-a\zeta)$ i.e. $a=1/2$ in
Sec.~\ref{gpdmodel}. A shift parameter of $a=1/2$ works well for NLO
but not for LO. In fact, the best description of the DVCS data in LO
is found for $a\simeq0$ for the MRST2001 LO PDF with $Q_0=1~\mbox{GeV}$ and
$\Lambda^{N_f=4,LO}_{QCD}=220~\mbox{MeV}$. Note that any $a\neq1/2$ will
violate the ``Munich symmetry'' of double distributions
\cite{rad2,much1} though still fullfilling all the other requirements.
Given the fact that the NLO parameterization fullfills all necessary
requirements and, save for absolute numbers, looks the same as the LO
parameterization, one can conclude that the LO parameterization is
still phenomenologically useful since it describes the data and can be
used for good quantitative estimates though it neglects higher order
corrections and violates some subtle symmetries.

\begin{figure}  
\mbox{\epsfig{file=h1wvsq.eps,width=8.5cm,height=10.5cm}} 
\vskip+0.2in
\mbox{\epsfig{file=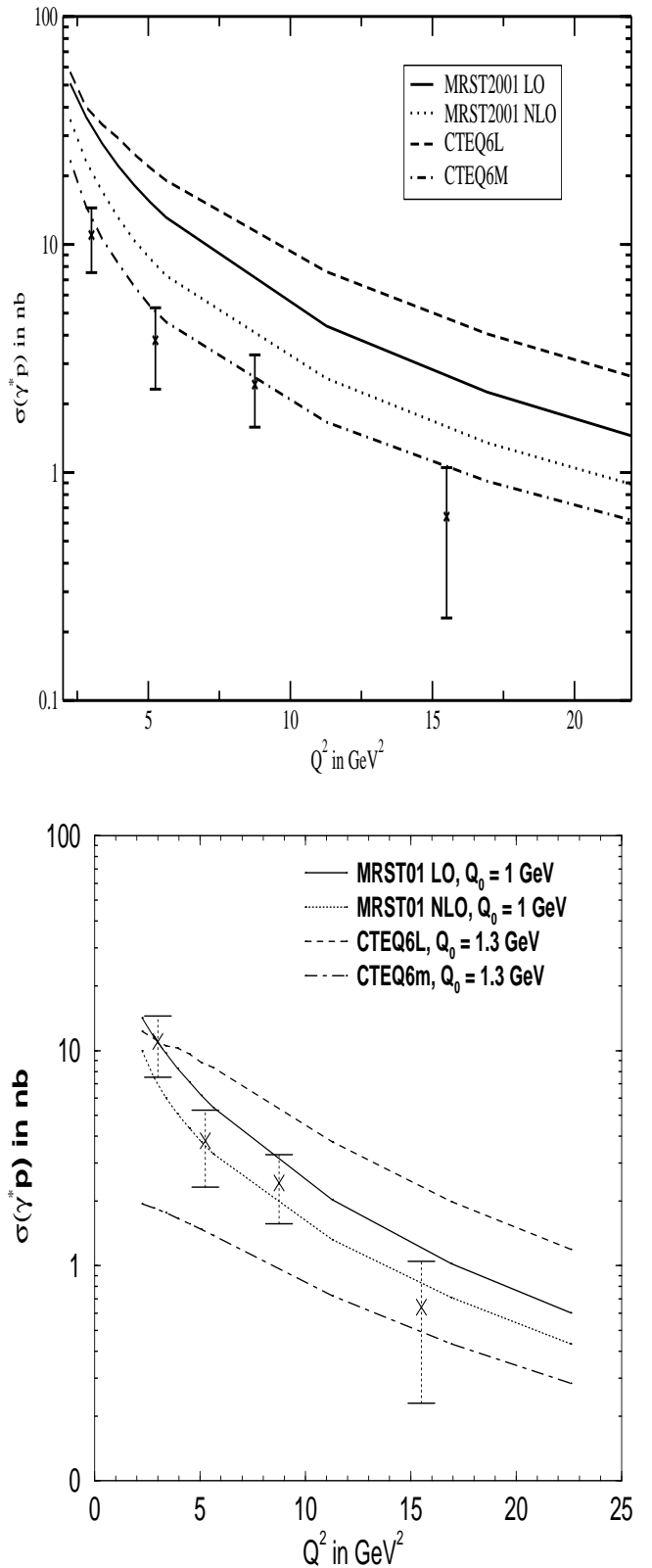,width=8.5cm,height=10.5cm}}
\caption{The photon level cross section, $\sigma (\gamma^* p \to \gamma p)$, in the average kinematics of the H1 data as a function of $Q^2$ at fixed $W=75~\mbox{GeV}$ with shift parameter $a=1/2$ (upper plot) and $a=0$ (lower plot).}
\label{h1wvsq}

\end{figure}

\begin{figure} 
\mbox{\epsfig{file=h1qvsw.eps,width=8.5cm,height=10.5cm}} 
\vskip+0.2in
\mbox{\epsfig{file=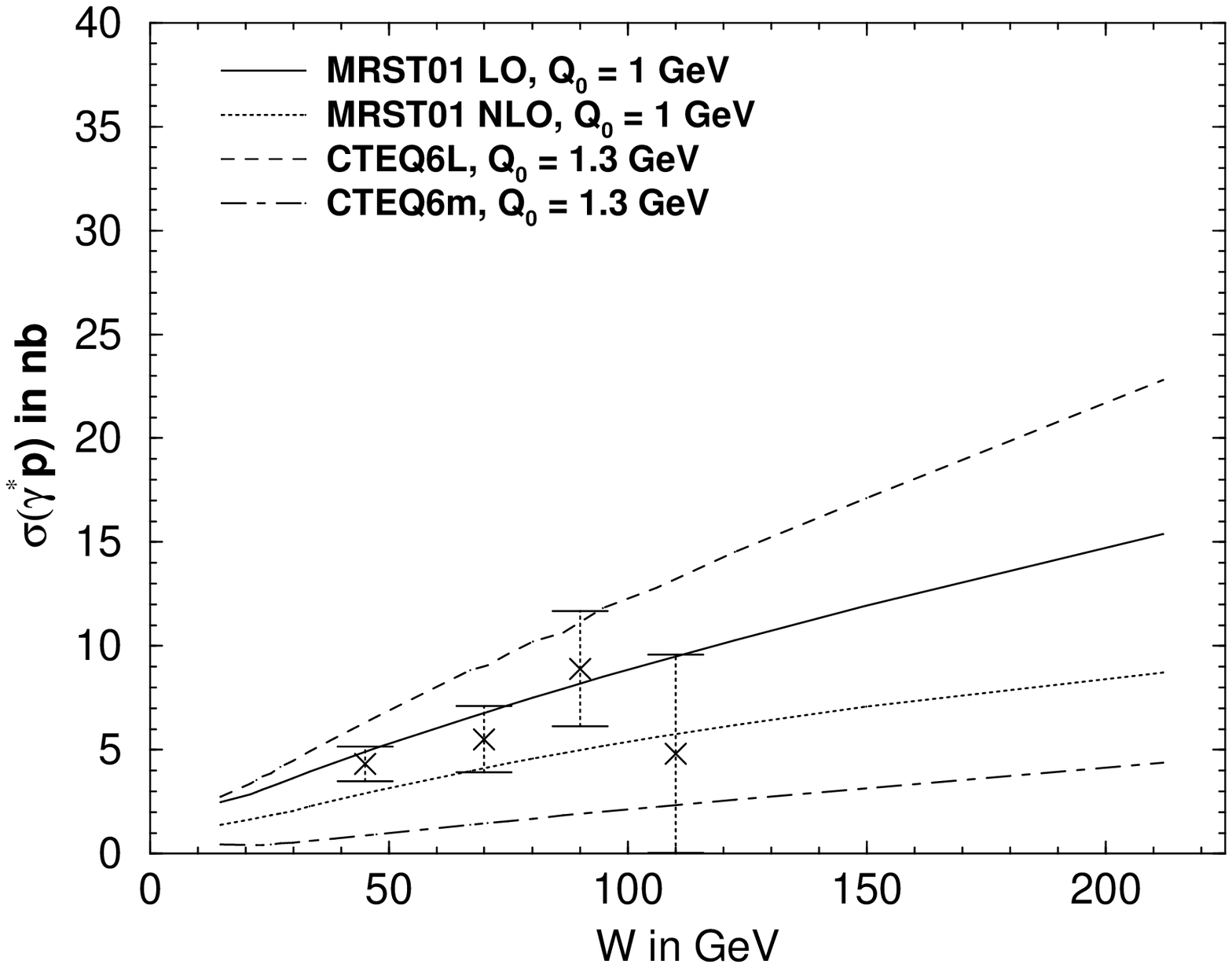,width=8.5cm,height=10.5cm}} 
\caption{The photon level cross section, $\sigma (\gamma^* p \to \gamma p)$, in the average kinematics of the H1 data as a function of $W$ at fixed $Q^2=4.5~\mbox{GeV}^2$ with shift parameter $a=1/2$ (upper plot) and $a=0$ (lower plot).}
\label{h1qvsw}
\end{figure}

Let me illustrate this with the example of the H1 data
Figs.~\ref{h1wvsq} and \ref{h1qvsw} for the DVCS $\gamma^*$-proton cross
section, $\sigma(\gamma^*p)$, Eq.~(\ref{sigonephot}). As I explained before, the
leading DVCS amplitude at small $\Bx$ is generated via the GPD $H$.
The interference term in the DVCS cross section is, after integration
over $\phi$, only a percent contribution to $\sigma(\gamma^*p)$. The BH term is
usually also negligible compared to the pure DVCS term, however, since
we are able to compute it unambiguously to high accuracy, one can
simply subtract the BH contribution from the data. In this case the
$t$ dependence can be simplified to an exponential form $e^{{\cal
    B}t}$. For the H1 data it is sufficient to take ${\cal B}$ to be a
constant. However, for the ZEUS data this is not sufficient anymore
(see \cite{afmmms}). Though not necessary, I will use the $Q^2$
dependent slope of \cite{afmmms}
\begin{equation}
{\cal B}(Q^2) = {\cal B}_0  \left(1 - C \ln \left(\frac{Q^2}{Q_0^2}\right)\right),  
\label{bq}
\end{equation}
with ${\cal B}_0 = 8~\mbox{GeV}^{-2}$, $Q_0 = 2~\mbox{GeV}^2$, $C =
0.15$. The reason for choosing such a parameterization are given in
great detail in \cite{afmmms} and need not be repeated here. A
physically intuitive explanation for this behavior of the slope is
given in \cite{afgpd3d}.  

As can be easily seen when comparing the upper and lower plots in
Figs.~\ref{h1wvsq} and \ref{h1qvsw}, the LO MRST2001 curve now compares
very favorably with the H1 data. In fact, it gives virtually the same
result as CTEQ6M i.e. it underestimates the ZEUS data somewhat.  For
the fixed target kinematics it is in agreement with the HERMES data on
the SSA ($0.21$) and CA ($0.09$) and the CLAS data ($0.17$) on the SSA
when kinematically averaged.

I would like to comment now on the $t$ dependence of GPD $H$ at $Q_0$.
Note that in the parameterization of the slope ${\cal B}$, I did not
introduce a $\Bx$ or $W$ dependence as is customarily done (see for
example \cite{ffs} and references therein) to account for cone
shrinkage i.e. the fact that the slope increases as $\Bx$ decreases
for constant $Q^2$. However the slope change in $\Bx$ for HERA
kinematics is only of the order of $10\%$ and can thus be neglected for
practical purposes. Furthermore, the necessity of a $Q^2$ dependent
slope signals a breakdown of factorizing the $t$ from the $X$ and $\zeta$
dependence as has always been done in modeling GPDs.  The breakdown at
small $\Bx$ does not occur until fairly large values of $Q^2$ which is
very suggestive of the following scenario: At the initial scale $Q_0$
one has a factorized component of the $t$ dependence which serves as a
normalization and will be different for valence-quarks, sea-quarks and
gluons. This difference in normalization between quarks and gluons
will change as the gluon mixes with the quark-singlet under
perturbative evolution.  This change in the normalization will be
$Q^2$ dependent and thus, the form of Eq.~(\ref{bq}) is very natural
since evolution resums logs of $Q^2$. The $\Bx$ dependence of the
slope can be generated similarly if the $\zeta$ dependence of the GPD is
$\zeta^{-\lambda-\alpha t}$ i.e. a regge-like dependence with $\alpha\simeq0.25~\mbox{GeV}^{-2}$
but possibly smaller, especially for gluons, which will also acquire a
logarithmic $Q^2$ dependence through evolution much like the
logarithmic slope of $F_2$ \cite{H1f2fit}.  This type of slope change
can also be parameterized with an exponential form as in Eq.~(\ref{bq})
with $Q^2$ replaced by $W$ i.e. $\Bx$. Note that in order to maintain
polynomiality of the GPD $H$ for $t\neq0$, the coefficients $A^i(\zeta)$ in
Eq.~(\ref{ajmerbl}) will also acquire a $t$ dependence in order to
compensate the extra factor of $\zeta^{-\alpha t}$.

A sensible parameterization for the $t$ dependence of the GPD $H$ at
$Q_0$ would thus be to choose a factorized exponential part with a
square root of the slope of $\frac{1}{2}{\cal B}_q\simeq 4-5~\mbox{GeV}^{-1}$ for the
quark sea and $\frac{1}{2}{\cal B}_g\simeq 2-2.5~\mbox{GeV}^{-1}$ for the gluon. The
valence quarks retain the dipole distribution in $t$ used in this
paper. The $t$ dependence in $\zeta$ can then be chosen to be $\zeta^{-\alpha t}$
for the quark sea and the gluon with $\alpha \simeq 0.25 ~\mbox{GeV}^{-2}$ for
simplicity. The valence distribution could in principle also have a
$\zeta^{-\alpha t}$ behavior but this will be extremely difficult to disentangle
from data from small $t$. A more accurate parameterization in $t$ will
require much more precise data as can be expected from the EIC.

This completes the phenomenological parameterization of the input GPD
$H$ in LO and NLO.

\section{Conclusions}
\label{conc}

I have given a detailed account of LO twist-3 effects in the WW
approximation including their perturbative evolution on DVCS
observables for kinematical settings equivalent to the HERA, EIC,
HERMES and CLAS experiments. Based on the successful GPD Ansatz of
\cite{afmmms}, I found that the twist-3 effects for the collider
settings are negligible save for the lowest values of $Q^2$ and $\Bx$.
For these $Q^2$ and $\Bx$ values the twist-3 effects still only reach
about $10\%$ in observables sensitive to the real part of DVCS
amplitudes namely the charge asymmetry and even less in observables
sensitive to the imaginary part of DVCS amplitudes such as the single
spin asymmetry. The twist-3 effects for the fixed target experiments
were only sizeable for the charge asymmetry at low $Q^2$ and $\Bx$,
however not larger than $10-15\%$ for the single spin asymmetry. The
common feature, of course, is the virtual disappearance of these
effects for $Q^2$ values larger than about $3-5~\mbox{GeV}^2$
depending on the value of $\Bx$. The relative smallness of twist-3
effects combined with the fact that twist-3 DVCS amplitudes in the WW
approximation are entirely expressible through twist-2 GPDs makes an
extraction of at least the unpolarized twist-2 GPD $H$ which is
leading in at least three of the four kinematical settings, entirely
feasible even with the current, relatively low statistics data. The
EIC with its high luminosity will then enable a high precision
extraction of the twist-2 GPD $H$.

Since the current data from H1, ZEUS and HERMES are already remarkably
restrictive for $H$, I give a first, phenomenological,
parameterization, though not a fit, in $X,\zeta$ and $t$ at a low
normalization point $Q_0$ which describes all available DVCS data from
HERA and HERMES in both LO and NLO and from CLAS in LO.

\section*{Acknowledgment}

This work was supported by the DFG under the Emmi-Noether grant
FR-1524/1-3. I would also like to acknowledge many useful discussions
with N.~Kivel, D.~M{\"u}ller, A.~Radyushkin, A.~Sch{\"a}fer, M.~Strikman and
C.~Weiss.


\begin{thebibliography}{99}
\bibitem{foot} Regular parton distributions are merely particle distributions not particle correlation functions.
\bibitem{footmom} The first moment counts the number of quarks in the
  proton and the second moment is a generalization of the momentum sum
  rule where the D-term generates all the deviation from unity as $\zeta$
  varies.
\end{thebibliography}
\end{document}